\documentclass[fleqn,10pt]{wlscirep}
\usepackage[utf8]{inputenc}
\usepackage[T1]{fontenc}

\usepackage{epsfig}
\usepackage{amssymb}
\usepackage{color,soul}
\usepackage[colorlinks=true]{hyperref}
\usepackage{pifont}
\usepackage[noabbrev]{cleveref}
\usepackage{booktabs}  % toprule, midrule, etc
\usepackage{multirow}
\usepackage{graphicx}
\usepackage{subfig}
\usepackage{todonotes}
\setuptodonotes{inline}

\newcommand{\anonymized}[1]{#1}

\newcommand{\beginsupplement}{%
    \setcounter{subsection}{0}
    \renewcommand{\thesubsection}{S\arabic{subsection}}%
    \setcounter{table}{0}
    \renewcommand{\thetable}{S\arabic{table}}%
    \setcounter{figure}{0}
    \renewcommand{\thefigure}{S\arabic{figure}}%
}

\title{End-To-End Prediction of Knee Osteoarthritis Progression With Multi-Modal Transformers}

\author[1,*]{Egor Panfilov}
\author[1,2]{Simo Saarakkala}
\author[1,2]{Miika T. Nieminen}
\author[1,3]{Aleksei Tiulpin}
\affil[1]{University of Oulu, Faculty of Medicine, Oulu, 90220, Finland}
\affil[2]{Oulu University Hospital, Department of Diagnostic Radiology, Oulu, 90220, Finland}
\affil[3]{Oulu University Hospital, Neurocenter, Oulu, 90220, Finland}
\affil[*]{corresponding author, egor.panfilov@oulu.fi}

\begin{abstract}
Knee Osteoarthritis (KOA) is a highly prevalent chronic musculoskeletal condition with no currently available treatment. The manifestation of KOA is heterogeneous and prediction of its progression is challenging. Current literature suggests that the use of multi-modal data and advanced modeling methods, such as the ones based on Deep Learning, has promise in tackling this challenge. To date, however, the evidence on the efficacy of this approach is limited. %However, Existing large-scale cohorts comprising multi-modal imaging data and modern computational methods pose an opportunity to clarify the overall value of imaging biomarkers in KOA progression prediction. 
In this study, we leveraged recent advances in Deep Learning and, using a Transformer approach, developed a unified framework for the multi-modal fusion of knee imaging data. Subsequently, we analyzed its performance across a range of scenarios by investigating multiple progression horizons -- from short-term to long-term. We report our findings using a large cohort (n=2421-3967) derived from the Osteoarthritis Initiative dataset. We show that structural knee MRI allows identifying radiographic KOA progressors on par with multi-modal fusion approaches, achieving an area under the ROC curve (ROC AUC) of 0.70-0.76 and Average Precision (AP) of 0.15-0.54 in 2-8 year horizons. Progression within 1 year was better predicted with a multi-modal method using X-ray, structural, and compositional MR images -- ROC AUC of 0.76(0.04), AP of 0.13(0.04) -- or via clinical data. Our follow-up analysis generally shows that prediction from the imaging data is more accurate for post-traumatic subjects, and we further investigate which subject subgroups may benefit the most. The present study provides novel insights into multi-modal imaging of KOA and brings a unified data-driven framework for studying its progression in an end-to-end manner, providing new tools for the design of more efficient clinical trials. The source code of our framework and the pre-trained models are made publicly available.

\end{abstract}

\begin{document}

\flushbottom
\maketitle

\thispagestyle{empty}

%%%%%%%%%%%%%%%%%%%%%%%%%%%%%%%%%%%%%%%%%%%%%%%%%%%%%%%%%%%%%%%%%%%%%%%%%%%%%%%%%%%%%%%%%%%%%%%%%%%%
\section*{Introduction}\label{sec:introduction}

\begin{figure}[ht]
\centering
\includegraphics[width=\linewidth]{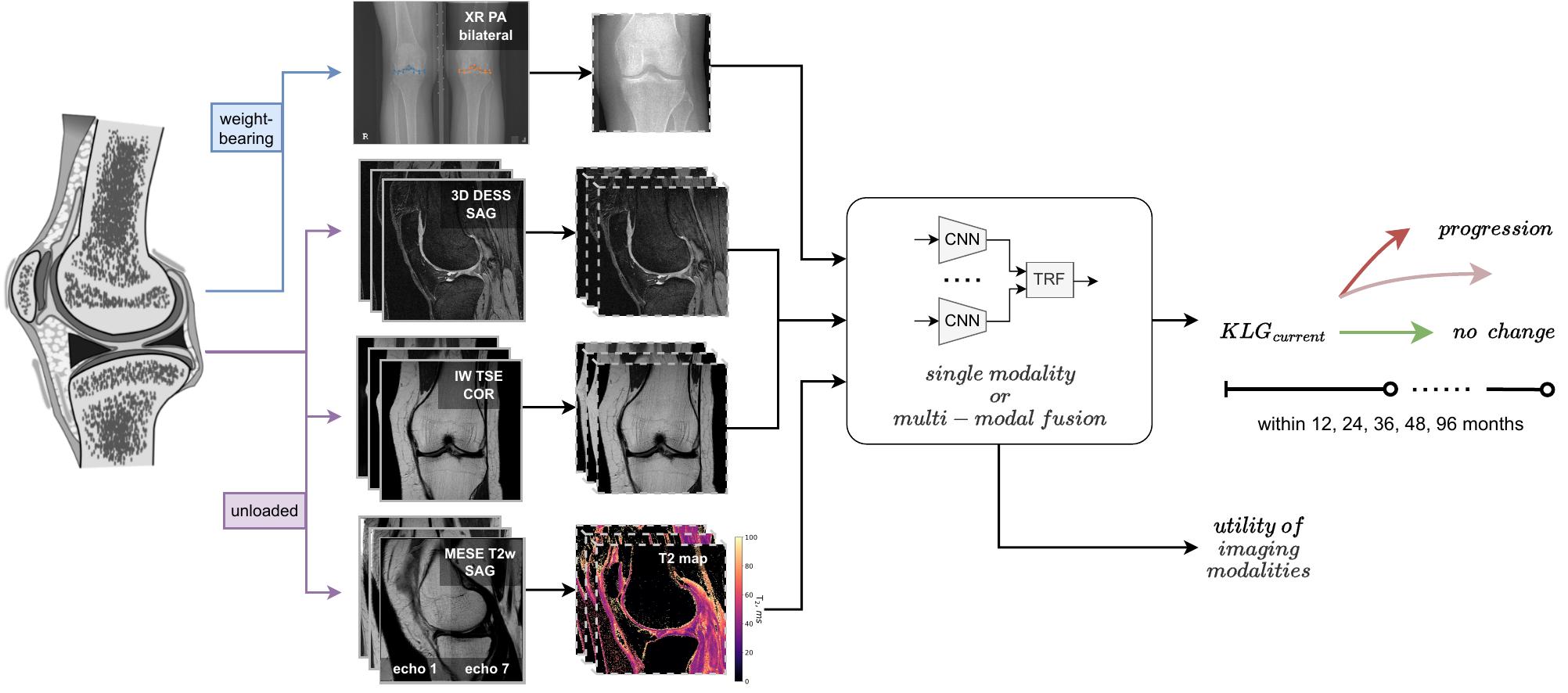}
\caption{Schematic overview of the proposed framework. Structural and compositional features imaged by several presumably complementary modalities are fused in a deep learning model, based on a composition of Convolutional Neural Networks (CNNs) and Transformers (TRFs). The imaging biomarkers are optimised to discriminate the knees that will undergo radiographic osteoarthritis (OA) progression versus the ones that will not progress within a certain time interval. Multiple intervals are considered to clarify the value of imaging modality in prediction of both rapid and slow OA progression. The models are additionally developed with each of the present modalities independently.}
\label{figure:pipeline}
\end{figure}

Knee osteoarthritis (KOA) is a chronic musculoskeletal disease affecting millions of people worldwide~\cite{cui2020global}. Progression of KOA results in degeneration of knee joint's bony and soft tissues, which is often accompanied by worsening in symptoms~\cite{neogi2009association}. Personalized prediction of structural KOA trajectory is important for multiple reasons, including early interventions and development of disease-modifying drugs, however, it is challenging due to high disease heterogeneity and rather poor understanding of KOA phenotypes~\cite{deveza2017knee,mobasheri2019recent,roemer2022structuralph}. %Two major variants of KOA are radiographic and symptomatic, with the former being considerably more prevalent~\cite{corti2003epidemiology,cui2020global}. 

Conventionally, the status of the suspected knees is assessed clinically from radiographic images. Weight-bearing X-ray images visualize alterations in bones' shape (e.g. osteophytes) and texture (e.g. subchondral sclerosis) with high contrast, as well as provide indirect measurements of cartilage and menisci degeneration via apparent joint space~\cite{hunter2006change}. These are the primary joint changes, and they are highly consistent across subjects with KOA. To date, the most established KOA severity scoring system -- Kellgren-Lawrence grading (KLG)\cite{kellgren1957radiological} -- is based on radiographic imaging.

Studies published during the past decade have shown that many soft tissue changes, e.g. in cartilage, menisci, ligaments, synovial and adipose tissues, are also associated with OA onset and progression~\cite{eckstein2015cartilage,ajuied2014anterior,kemnitz2018role,wang2019knee,wang2019quantitative}. They are not visible in radiographs but can be detected and tracked using Magnetic Resonance Imaging (MRI), which enables three-dimensional imaging of the joint. Knee MRI studies typically include several MR imaging protocols with complementary contrasts, and they target morphological factors in major joint tissues, such as the severity of osteophytes, cartilage thickness, and meniscal and ligament tears. The MRI protocols can be divided into structural -- targetting tissue morphology -- and compositional MRI -- reflecting microstructure and biochemical content. The most apparent morphological changes in soft tissues have been incorporated into advanced grading schemes, such as MOAKS~\cite{hunter2011evolution}, however, utilization of such schemes for studying KOA progression remains limited~\cite{namiri2021deep,roemer2022structuralti}. Quantitative MRI (qMRI) protocols, such as $T_2$-mapping, have been getting increased attention due to their sensitivity to compositional tissue changes (e.g. collagen anisotropy in cartilage and meniscus in early KOA~\cite{prasad2013t1rho,hunter2015oarsi,emanuel2021relation,zarins2010cartilage}, fatty infiltration of muscles~\cite{eck2023quantitative}) and considerable technology readiness level~\cite{hilbert2018accelerated,chalian2021qiba}. Overall, despite the rich information provided by multi-sequence MRI in addition to radiography and sensitivity to early tissue changes, the real prognostic utility of MRI and, specifically, qMRI in KOA remains understudied~\cite{razmjoo2021t2}.

The vast majority of prior art on MRI in KOA progression prediction operated with limited sample sizes and highly interpretable and localized imaging biomarkers, which are typically extracted via image segmentation and basic radiomics~\cite{eckstein2006proposal,neogi2013magnetic}. Such conventional biomarkers are designed using a "bottom-up" approach, primarily describing apparent changes that occur in major joint tissues, particularly, in cartilage. As a result, the role of less affected tissues remains unstudied, and it is gaining attention only recently~\cite{kemnitz2018role,wang2019knee,li2022mri}. Another limitation of many prior works is that they perform aggressive subject exclusion for the definition of groups, omitting the study participants with mixed and inconsistent findings. This process allows studying the sensitivity of developed biomarkers in the discrimination of small-scale groups, while severely compromising/underestimating their specificity (i.e. generalization)~\cite{iriondo2021towards}. While this knowledge lays the foundation for clinical management of KOA subjects by fine-grained differentiation of disease progression, it does not necessarily answer the question of how the disease will progress in the future in a particular subject from a general population.

Modern computational methods, such as the ones based on Deep Learning (DL), have made possible the analysis of large-scale imaging studies and the development of new personalized prediction models~\cite{litjens2017survey,miotto2018deep}. With DL, the design of imaging biomarkers can be seen as "top-down" process. Here, the informative features that are discriminative w.r.t. the defined target are first automatically derived in a data-driven manner~\cite{rueckert2019model}. Subsequently, the learned features and their interaction are analyzed from the model by factorization of model activations into interpretable concepts defined by a human expert. While interpretability of DL models remains challenging~\cite{jin2021one,han2022explanation}, such methods allow to understand the peak performance of certain data in the considered task, long before the clinically applicable biomarkers are designed~\cite{hirvasniemi2022knee}.

In the KOA domain, Tiulpin et al~\cite{tiulpin2019multimodal} have previously shown superior performance of DL applied to raw radiographic images in comparison to demographic variables and gold-standard KLG in the task of radiographic progression prediction. Studies on MRI data analysis in this scope, however, are very sparse. Wang et al~\cite{wang2019total} demonstrated high performance of DL with two MRI protocols in predicting whether the knee will undergo total knee replacement (TKR) within 9 years from the exam. In the same problem, but at 5 years horizon, Tolpadi et al~\cite{tolpadi2020deep} contrasted radiographic and MR images showing a slight advantage of the latter modality. While TKR is regulatory-approved as a KOA endpoint, it is not inherent to the disease, and we argue that it is a noisy progression surrogate. To this end, the recent work of Panfilov et al~\cite{panfilov2022predicting} have recently compared X-ray images and structural MRI in the prediction of radiographic KOA progression (increase of KLG as in the work of Tiulpin~\cite{tiulpin2019multimodal} et al.) within 8 years. All in all, KOA forecasting over short-term, which is more valuable for clinical trials, has not been thoroughly addressed. On top of that, the complementary value of clinically accessible imaging modalities, especially, compositional MRI, in identification of progressors remains an open question.

To date, the majority of DL-based multi-modal methods in medical image computing either perform aggressive data dimensionality reduction~\cite{banerjee2018transfer,huang2020fusion} or multi-stage late fusion~\cite{tiulpin2019multimodal,hu2020deep}, where the modalities are first processed separately and then combined in a second-level shallow model. Both considerations are applied due to high memory demand in processing typically large medical images.
Accordingly, both of the aforementioned techniques limit the model's capabilities to derive rich and interrelated features. Lately, thanks to advances in computational platforms and DL methods, unified attention-based methods, such as Transformers~\cite{tsai2019multimodal,dosovitskiy2020image}, were developed. Transformers have opened a possibility for holistic modeling in diverse multi-modal scenarios, with little to no modification of the original data~\cite{tulder2021multi,nguyen2022climat}. In medical imaging, they were shown to often provide higher accuracy, particularly, when used with pre-training or in high volume data setting~\cite{ascoli2021convit,jun2021medical}.

In this study, we introduce a multi-modal DL-based method for predicting radiographic KOA progression (hereinafter referred to as "KOA progression") and investigate the value of various modalities in this task. The contributions of our work are three-fold:
\begin{itemize}
    \item We propose a new end-to-end method to study KOA progression from multi-modal imaging data. We apply the method for prediction of rapid, middle-, and long-term radiographic progression, where we clarify the predictive value of imaging in the task and establish the new baseline models.
    \item We comprehensively analyze the complementary value of common imaging modalities (X-ray, structural, and compositional MRI) with respect to the considered outcomes. Our study is the first to use the quantitative $T_2$ maps of MRI in an end-to-end predictive model, and among the few to study compositional MRI in a large-scale setting.
    \item We analyze the efficacy of the best-performing models across different subject sub-groups and discuss the directions for further development of top-down methods for KOA progression prediction.
\end{itemize}

%%%%%%%%%%%%%%%%%%%%%%%%%%%%%%%%%%%%%%%%%%%%%%%%%%%%%%%%%%%%%%%%%%%%%%%%%%%%%%%%%%%%%%%%%%%%%%%%%%%%
\section*{Results}\label{sec:results}
\subsection*{Training and testing datasets}
Five observation intervals were considered (0-12/24/36/48/96 months) to derive 5 independent datasets from the Osteoarthritis Initiative (OAI) database. The complete sample selection procedure is presented in Figure~\ref{figure:selection}. The most common reasons for exclusion were patient dropouts and missing clinical or imaging data. The progression target was defined based on the change in KLG within the considered interval. The knees with no recorded change in KLG were assigned to the "control" group and the ones with observed worsening of KLG - to the "progressor" group. Following the popular research practice, grades KLG0 and KLG1 were pooled together, as the corresponding change is often not considered clinically significant or reliable (KL1 is defined as ``doubtful OA'')~\cite{riddle2016incident,tiulpin2019multimodal,guan2022deep}. After grade pooling, a small number of subjects still showed an improvement in KLG, with or without accompanying worsening. To avoid ambiguity in the definition of disease progression, those subjects were excluded from the study. The final sample sizes were 3967, 3735, 3585, 3448, and 2421 for 12m, 24m, 36m, 48m, and 96m intervals, respectively. The ratio of progressors to the total number of subjects was notably higher with longer observation periods - 5.7, 8.4, 11.9, 14.5, and 27.7\% for 12m, 24m, 36m, 48m, and 96m, respectively.

The resulting datasets were split into training, validation, and testing subsets. In the OAI, the subjects were observed at multiple data acquisition sites. All the subjects from the site "D" were assigned to the test set. While the acquisition protocols in the OAI are supposed to be standardized between the sites, a small domain shift between the images from different sites is still present. This subject allocation scheme allowed us to additionally model the potential discrepancy between training-time and testing-time images and, thus, make the evaluation more objective. The testing subsets' sample sizes were 1016, 933, 896, 867, and 626 for 12m, 24m, 36m, 48m, and 96m targets, respectively, which is 25-26\% of the total sample. The remaining samples were split following a 5-fold stratifed cross-validation scheme ($\approx$80/20\%) while balancing the ratio of controls and progressors in the training and the validation subsets for each split (no overlapping subject-wise between the training and validation sets).

\begin{table}[ht]
\centering
\caption{Description of the created datasets. Only one knee per subject was selected. The values for Age and BMI represent variable mean and standard deviation.}
\label{table:demogr}
\begin{tabular}{l|c|c|c|c|c}
\toprule
\multicolumn{1}{c|}{\textbf{Target} $>$} & \textbf{12m} & \textbf{24m} & \textbf{36m} & \textbf{48m} & \textbf{96m} \\ \hline
\# subjects/knees, of which: & 3967 & 3735 & 3585 & 3448 & 2421 \\
\textit{- controls} & 3740 & 3420 & 3160 & 2947 & 1751 \\
\textit{- progressors} & 227 & 315 & 425 & 501 & 670 \\ \midrule
Age & 61.1~(9.2) & 61.1~(9.1) & 60.9~(9.1) & 61.0~(9.1) & 60.1~(8.8) \\
BMI & 28.5~(4.8) & 28.4~(4.7) & 28.4~(4.7) & 28.4~(4.7) & 28.1~(4.6) \\
Sex: \textit{F, M} & 2314, 1653 & 2176, 1559 & 2091, 1494 & 2007, 1441 & 1400, 1021 \\ \midrule
WOMAC: \textit{{[}0-10{]}, (10-100{]}} & 2534, 1433 & 2401, 1334 & 2309, 1276 & 2237, 1211 & 1644, 777 \\
\begin{tabular}[c]{@{}c@{}}KLG: \textit{0, 1, 2, 3}\\ \phantom{x} \end{tabular} & \begin{tabular}[c]{@{}c@{}}1555, 750,\\ 1113, 549\end{tabular} & \begin{tabular}[c]{@{}c@{}}1454, 711,\\ 1057, 513\end{tabular} & \begin{tabular}[c]{@{}c@{}}1403, 683,\\ 1016, 483\end{tabular} & \begin{tabular}[c]{@{}c@{}}1354, 672,\\ 983, 439\end{tabular} & \begin{tabular}[c]{@{}c@{}}1154, 576,\\ 479, 212\end{tabular} \\
Prior injury: \textit{no, yes} & 2907, 1060 & 2722, 1013 & 2613, 972 & 2527, 921 & 1782, 639 \\
Prior surgery: \textit{no, yes} & 3543, 424 & 3322, 413 & 3209, 376 & 3087, 361 & 2189, 232\\ \bottomrule
\end{tabular}
\end{table}

\subsection*{Progression prediction from individual modalities}
\paragraph{Clinical data and semi-quantitative X-ray assessments}
To better understand the predictive power of common clinical risk factors, a set of baseline models was developed. The variables included subject age, sex, BMI, history of past surgeries and injuries, symptomatic score (WOMAC; Western Ontario and McMaster Universities Osteoarthritis Index)~\cite{bellamy1988validation}, as well as routinely assessed radiographic KLG score. The models along with their performance are described in Table \ref{table:clin}. For the 12-month prediction horizon, adding WOMAC score and history of knee alterations yielded a notable increase of 0.07 in both average ROC AUC (\textit{p=0.079}) and AP (\textit{p=0.042}). Inclusion of KLG further improved AP by 0.03 (\textit{p=0.030}), suggesting an added value of imaging in predicting progression short-term. For 24m-48m horizons, similar findings were observed, however, the predictive power of knee history and WOMAC score decreased, and the additional value of KLG, given other risk factors, was marginal. Interestingly, for 96m horizon, the presence of knee alteration history, WOMAC (model \textit{C3}), and also KLG (model \textit{C4}) yielded a notable increase in ROC AUC (0.03 [\textit{p=0.031}] and 0.05 [\textit{p=0.008}], respectively) and AP (0.05 [\textit{p=0.019}] and 0.05 [\textit{p=0.023}], respectively). Towards longer horizons, the average performance of all models grew faster in AP than the prevalence rate, suggesting that the identification of long-term progressors compared to rapid ones is more feasible. Taking the observed performance benefits of KLG into account, a purely non-imaging model \textit{C3} was used as a baseline in subsequent analysis.

\begin{table}[ht]
\centering
\caption{Performance of the models based on the widely accessible clinical data. The variables include body mass index (BMI), radiographic severity via Kellgren-Lawrence grade (KLG), history of past injuries (Inj) and surgeries (Surg), symptomatic and knee function assessment via Western Ontario and McMaster Universities Arthritis Index (WOMAC). Prevalence indicates the rate of the progressed knees and, accordingly, a performance of a naive classifier. The values show average precision (AP) and area under the ROC curve (ROC AUC) along with the standard errors.}
\label{table:clin}
\begin{tabular}{@{}c|l|c|c|c|c|c@{}}
\toprule
 & \multicolumn{1}{c|}{\textbf{Target} $>$} & \textbf{12m} & \textbf{24m} & \textbf{36m} & \textbf{48m} & \textbf{96m} \\
\midrule
\textbf{Model} & \multicolumn{1}{c}{\textbf{Data}} & \multicolumn{5}{|c}{\textbf{ROC AUC @ target}} \\
\midrule
C1 & age, sex, BMI & 0.64$_{\pm 0.04}$ & 0.62$_{\pm 0.04}$ & 0.63$_{\pm 0.03}$ & 0.65$_{\pm 0.03}$ & 0.67$_{\pm 0.02}$ \\
C2 & + KLG & 0.63$_{\pm 0.04}$ & 0.65$_{\pm 0.03}$ & 0.66$_{\pm 0.03}$ & 0.68$_{\pm 0.03}$ & 0.74$_{\pm 0.02}$ \\
C3 & + Surg, Inj, WOMAC & 0.71$_{\pm 0.04}$ & 0.69$_{\pm 0.04}$ & 0.66$_{\pm 0.03}$ & 0.68$_{\pm 0.03}$ & 0.70$_{\pm 0.02}$ \\
C4 & + Surg, Inj, WOMAC, KLG & 0.72$_{\pm 0.04}$ & 0.70$_{\pm 0.03}$ & 0.68$_{\pm 0.03}$ & 0.70$_{\pm 0.03}$ & 0.75$_{\pm 0.02}$ \\ \midrule
\textbf{Model} & \multicolumn{1}{c}{\textbf{Data}} & \multicolumn{5}{|c}{\textbf{AP @ target}} \\
\midrule
C1 & age, sex, BMI & 0.06$_{\pm 0.01}$ & 0.08$_{\pm 0.01}$ & 0.17$_{\pm 0.02}$ & 0.19$_{\pm 0.02}$ & 0.36$_{\pm 0.03}$ \\
C2 & + KLG & 0.06$_{\pm 0.01}$ & 0.09$_{\pm 0.01}$ & 0.17$_{\pm 0.02}$ & 0.20$_{\pm 0.02}$ & 0.42$_{\pm 0.03}$ \\
C3 & + Surg, Inj, WOMAC & 0.13$_{\pm 0.04}$ & 0.16$_{\pm 0.04}$ & 0.20$_{\pm 0.03}$ & 0.22$_{\pm 0.03}$ & 0.41$_{\pm 0.03}$ \\
C4 & + Surg, Inj, WOMAC, KLG & 0.16$_{\pm 0.05}$ & 0.17$_{\pm 0.04}$ & 0.20$_{\pm 0.03}$ & 0.23$_{\pm 0.03}$ & 0.46$_{\pm 0.03}$ \\
 & \textit{prevalence} & \textit{0.04} & \textit{0.06} & \textit{0.10} & \textit{0.13} & \textit{0.25} \\
\bottomrule
\end{tabular}
\end{table}

\paragraph{Raw X-ray images}
End-to-end models trained on raw radiographic images (XR) showed moderate performance at all horizons, as summarized in Table~\ref{table:singlemod}. Compared to the baseline, the modesl were inferior in both metrics at 12m and comparable at 24m. From 36m onwards, the models showed higher scores than the baseline, reaching statistically significant (\textit{p<0.021}) improvements of 0.08 in AP for 48-96m targets.

\paragraph{MRI data}
The performance of MRI-based models varied depending on whether structural (DESS/TSE) or compositional ($T_2$map) protocol was used (see Table~\ref{table:singlemod}). Structural modalities showed improved performance in ROC AUC - comparable to \textit{C3} and higher than \textit{X} at 12m, and generally higher than both from 24m onward. Most notable increases in average AUC were observed for the 24m and 96m horizons. $T_2$map-based model \textit{M3} showed similar ROC AUCs as the XR one. In terms of AP, all models were similar to \textit{X}, except for 48m (where the scores were marginally lower by 0.02-0.03) and 96m (where they improved the mean score by notable 0.05-0.07). Of all the observed improvements, the significant ones were found mostly for 96m prediction horizon. Here, all the MRI models were significantly better than the clinical baseline in both metrics (\textit{p<0.023}). When compared against XR, the structural MRI protocols (\textit{M1} [DESS] and \textit{M2} [TSE]) also showed higher performance, both in ROC AUC (\textit{p=0.020} and \textit{p=0.007}, respectively) and AP (\textit{p=0.138} and \textit{p=0.017}, respectively). The model \textit{M1} was significantly better than the clinical baseline in ROC AUC also at 48m (\textit{p=0.030}).

\begin{table}[ht]
\centering
\caption{Performance of the single modality models, based on X-ray (XR) or MRI (DESS, TSE, T$_2$\text{map}). The best non-imaging clinical model is chosen for reference. The values show average precision (AP) and area under the ROC curve (ROC AUC) along with the standard errors. Statistically significant improvements (\textit{p<0.050}) are marked with superscripts: \texttt{c} -- vs. C3, \texttt{x} -- vs. X.}
\label{table:singlemod}
\begin{tabular}{c|l|c|rrrrr}
\toprule
 & \multicolumn{1}{c|}{\textbf{}} & \multicolumn{1}{l|}{\textbf{Target $>$}} & \multicolumn{1}{c|}{\textbf{12m}} & \multicolumn{1}{c|}{\textbf{24m}} & \multicolumn{1}{c|}{\textbf{36m}} & \multicolumn{1}{c|}{\textbf{48m}} & \multicolumn{1}{c}{\textbf{96m}} \\
\midrule
 \textbf{Model} & \multicolumn{1}{c|}{\textbf{Data}} & \textbf{Arch} & \multicolumn{5}{c}{\textbf{ROC AUC @ target}} \\ \midrule
C3 & \begin{tabular}[c]{@{}l@{}}age, sex, BMI, Surg,\\ Inj, WOMAC\end{tabular} & LR & \multicolumn{1}{r|}{0.71$_{\pm 0.04}$} & \multicolumn{1}{r|}{0.69$_{\pm 0.04}$} & \multicolumn{1}{r|}{0.66$_{\pm 0.03}$} & \multicolumn{1}{r|}{0.68$_{\pm 0.03}$} & 0.70$_{\pm 0.02}$ \\
X & XR & XR1 & \multicolumn{1}{r|}{0.65$_{\pm 0.04}$} & \multicolumn{1}{r|}{0.68$_{\pm 0.04}$} & \multicolumn{1}{r|}{0.69$_{\pm 0.03}$} & \multicolumn{1}{r|}{0.70$_{\pm 0.03}$} & 0.73$_{\pm 0.02}$ \\
M1 & DESS & MR1 & \multicolumn{1}{r|}{0.71$_{\pm 0.04}$} & \multicolumn{1}{r|}{0.74$_{\pm 0.03}$} & \multicolumn{1}{r|}{0.70$_{\pm 0.03}$} & \multicolumn{1}{r|}{0.73$_{\pm 0.03}^{\texttt{(c)}}$} & 0.76$_{\pm 0.02}^{\texttt{(cx)}}$ \\
M2 & TSE & MR1 & \multicolumn{1}{r|}{0.71$_{\pm 0.04}$} & \multicolumn{1}{r|}{0.75$_{\pm 0.03}$} & \multicolumn{1}{r|}{0.70$_{\pm 0.03}$} & \multicolumn{1}{r|}{0.70$_{\pm 0.03}$} & 0.78$_{\pm 0.02}^{\texttt{(cx)}}$ \\
M3 & T$_2$\text{map} & MR1 & \multicolumn{1}{r|}{0.69$_{\pm 0.04}$} & \multicolumn{1}{r|}{0.69$_{\pm 0.04}$} & \multicolumn{1}{r|}{0.68$_{\pm 0.03}$} & \multicolumn{1}{r|}{0.68$_{\pm 0.03}$} & 0.74$_{\pm 0.02}^{\texttt{(c)}}$ \\
\midrule
\textbf{Model} & \multicolumn{1}{c|}{\textbf{Data}} & \textbf{Arch} & \multicolumn{5}{c}{\textbf{AP @ target}} \\
\midrule
C3 & \begin{tabular}[c]{@{}l@{}}age, sex, BMI, Surg,\\ Inj, WOMAC\end{tabular} & LR & \multicolumn{1}{r|}{0.13$_{\pm 0.04}$} & \multicolumn{1}{r|}{0.16$_{\pm 0.04}$} & \multicolumn{1}{r|}{0.20$_{\pm 0.03}$} & \multicolumn{1}{r|}{0.22$_{\pm 0.03}$} & 0.41$_{\pm 0.03}$ \\
X & XR & XR1 & \multicolumn{1}{r|}{0.10$_{\pm 0.03}$} & \multicolumn{1}{r|}{0.15$_{\pm 0.04}$} & \multicolumn{1}{r|}{0.23$_{\pm 0.03}$} & \multicolumn{1}{r|}{0.30$_{\pm 0.04}^{\texttt{(c)}}$} & 0.49$_{\pm 0.03}^{\texttt{(c)}}$ \\
M1 & DESS & MR1 & \multicolumn{1}{r|}{0.12$_{\pm 0.04}$} & \multicolumn{1}{r|}{0.15$_{\pm 0.03}$} & \multicolumn{1}{r|}{0.24$_{\pm 0.04}$} & \multicolumn{1}{r|}{0.27$_{\pm 0.03}$} & 0.54$_{\pm 0.03}^{\texttt{(c)}}$ \\
M2 & TSE & MR1 & \multicolumn{1}{r|}{0.09$_{\pm 0.02}$} & \multicolumn{1}{r|}{0.16$_{\pm 0.04}$} & \multicolumn{1}{r|}{0.21$_{\pm 0.03}$} & \multicolumn{1}{r|}{0.27$_{\pm 0.03}$} & 0.56$_{\pm 0.03}^{\texttt{(cx)}}$ \\
M3 & T$_{2}$\text{map} & MR1 & \multicolumn{1}{r|}{0.11$_{\pm 0.04}$} & \multicolumn{1}{r|}{0.16$_{\pm 0.04}$} & \multicolumn{1}{r|}{0.24$_{\pm 0.04}$} & \multicolumn{1}{r|}{0.28$_{\pm 0.04}$} & 0.54$_{\pm 0.03}^{\texttt{(c)}}$ \\
 & \textit{prevalence} &  & \multicolumn{1}{c|}{\textit{0.04}} & \multicolumn{1}{c|}{\textit{0.06}} & \multicolumn{1}{c|}{\textit{0.10}} & \multicolumn{1}{c|}{\textit{0.13}} & \multicolumn{1}{c}{\textit{0.25}} \\
\bottomrule
\end{tabular}
\end{table}

\subsection*{Multi-modal fusion}
To clarify the complementary value of the considered imaging modalities, we performed an exhaustive experimental investigation. Here, three sets of models were developed based on the individual modalities studied earlier: fusion of XR with single MRI protocol (XR1MR1), two MRI protocols (MR2), and XR with two MRI protocols (XR1MR2). The best models selected within each setting are summarized in Table~\ref{table:fusion_best_stat} and the complete results including all models can be found in Table~\ref{table:fusion}.

\begin{table}[ht]
\centering
\caption{Performance of the selected top performing fusion models. Model F1 combines X-ray with single MRI sequence, F4-F5 - two MRI sequences, F8 - X-ray with two MRI sequences. The values show average precision (AP) and area under the ROC curve (ROC AUC) along with the standard errors. Statistically significant improvements (\textit{p<0.050}) are marked with superscripts: \texttt{c} -- vs. C3 (clinical), \texttt{x} -- vs. X (XR), \texttt{m} -- vs. M1 (DESS). The extended version of this table showing full factorial analysis of modality fusion can be found in Supplemental Table~\ref{table:fusion}. More details on the architecture of the fusion models are provided in Supplemental~\ref{figure:archs}.}
\label{table:fusion_best_stat}
\begin{tabular}{c|l|c|rrrrr}
\toprule
 & \multicolumn{1}{c|}{\textbf{}} & \textbf{Target $>$} & \multicolumn{1}{c|}{\textbf{12m}} & \multicolumn{1}{c|}{\textbf{24m}} & \multicolumn{1}{c|}{\textbf{36m}} & \multicolumn{1}{c|}{\textbf{48m}} & \multicolumn{1}{c}{\textbf{96m}} \\ \midrule
\textbf{Model} & \multicolumn{1}{c|}{\textbf{Data}} & \textbf{Arch} & \multicolumn{5}{c}{\textbf{ROC AUC @ target}} \\
\midrule
F1 & XR, DESS & XR1MR1 & \multicolumn{1}{r|}{0.76$_{\pm 0.03}^{\texttt{(x)}}$} & \multicolumn{1}{r|}{0.72$_{\pm 0.04}$} & \multicolumn{1}{r|}{0.70$_{\pm 0.03}$} & \multicolumn{1}{r|}{0.74$_{\pm 0.03}^{\texttt{(c)}}$} & 0.77$_{\pm 0.02}^{\texttt{(cx)}}$ \\
F4 & DESS, TSE & MR2 & \multicolumn{1}{r|}{0.74$_{\pm 0.03}$} & \multicolumn{1}{r|}{0.74$_{\pm 0.03}$} & \multicolumn{1}{r|}{0.70$_{\pm 0.03}$} & \multicolumn{1}{r|}{0.71$_{\pm 0.03}$} & 0.76$_{\pm 0.02}^{\texttt{(cx)}}$ \\
F5 & DESS, T$_2$\text{map} & MR2 & \multicolumn{1}{r|}{0.72$_{\pm 0.04}$} & \multicolumn{1}{r|}{0.74$_{\pm 0.03}$} & \multicolumn{1}{r|}{0.72$_{\pm 0.03}^{\texttt{(c)}}$} & \multicolumn{1}{r|}{0.73$_{\pm 0.03}^{\texttt{(c)}}$} & 0.76$_{\pm 0.02}^{\texttt{(c)}}$ \\
F8 & XR, DESS, T$_2$\text{map} & XR1MR2 & \multicolumn{1}{r|}{0.76$_{\pm 0.04}^{\texttt{(xm)}}$} & \multicolumn{1}{r|}{0.75$_{\pm 0.03}^{\texttt{(c)}}$} & \multicolumn{1}{r|}{0.70$_{\pm 0.03}$} & \multicolumn{1}{r|}{0.73$_{\pm 0.03}^{\texttt{(c)}}$} & 0.77$_{\pm 0.02}^{\texttt{(cx)}}$ \\
U & \textit{F8 \& C3 vars} & XR1MR2C1 & \multicolumn{1}{r|}{0.71$_{\pm 0.04}$} & \multicolumn{1}{r|}{0.73$_{\pm 0.03}$} & \multicolumn{1}{r|}{0.70$_{\pm 0.03}$} & \multicolumn{1}{r|}{0.72$_{\pm 0.03}^{\texttt{(c)}}$} & \multicolumn{1}{r}{0.76$_{\pm 0.02}^{\texttt{(cx)}}$} \\
\midrule
\textbf{Model} & \multicolumn{1}{c|}{\textbf{Data}} & \textbf{Arch} & \multicolumn{5}{c}{\textbf{AP @ target}} \\
\midrule
F1 & XR, DESS & XR1MR1 & \multicolumn{1}{r|}{0.11$_{\pm 0.03}$} & \multicolumn{1}{r|}{0.15$_{\pm 0.03}$} & \multicolumn{1}{r|}{0.23$_{\pm 0.03}$} & \multicolumn{1}{r|}{0.29$_{\pm 0.04}^{\texttt{(c)}}$} & 0.56$_{\pm 0.03}^{\texttt{(cx)}}$ \\
F4 & DESS, TSE & MR2 & \multicolumn{1}{r|}{0.12$_{\pm 0.03}$} & \multicolumn{1}{r|}{0.16$_{\pm 0.03}$} & \multicolumn{1}{r|}{0.20$_{\pm 0.03}$} & \multicolumn{1}{r|}{0.27$_{\pm 0.03}$} & 0.55$_{\pm 0.03}^{\texttt{(c)}}$ \\
F5 & DESS, T$_2$\text{map} & MR2 & \multicolumn{1}{r|}{0.12$_{\pm 0.04}$} & \multicolumn{1}{r|}{0.16$_{\pm 0.03}$} & \multicolumn{1}{r|}{0.24$_{\pm 0.03}$} & \multicolumn{1}{r|}{0.28$_{\pm 0.03}$} & 0.54$_{\pm 0.03}^{\texttt{(c)}}$ \\
F8 & XR, DESS, T$_2$\text{map} & XR1MR2 & \multicolumn{1}{r|}{0.13$_{\pm 0.04}$} & \multicolumn{1}{r|}{0.16$_{\pm 0.03}$} & \multicolumn{1}{r|}{0.22$_{\pm 0.03}$} & \multicolumn{1}{r|}{0.27$_{\pm 0.03}$} & 0.57$_{\pm 0.03}^{\texttt{(cx)}}$ \\
U & \textit{F8 \& C3 vars} & XR1MR2C1 & \multicolumn{1}{r|}{0.10$_{\pm 0.02}$} & \multicolumn{1}{r|}{0.15$_{\pm 0.03}$} & \multicolumn{1}{r|}{0.23$_{\pm 0.03}$} & \multicolumn{1}{r|}{0.26$_{\pm 0.03}$} & \multicolumn{1}{r}{0.55$_{\pm 0.03}^{\texttt{(c)}}$} \\
 & \textit{prevalence} &  & \multicolumn{1}{c|}{\textit{0.04}} & \multicolumn{1}{c|}{\textit{0.06}} & \multicolumn{1}{c|}{\textit{0.10}} & \multicolumn{1}{c|}{\textit{0.13}} & \multicolumn{1}{c}{\textit{0.25}} \\
\bottomrule
\end{tabular}
\end{table}

\paragraph{Fusion of MRI sequences}
A combination of two MRI modalities resulted in only marginal improvement over individual structural MR sequences. Particularly, the fusion of DESS and TSE showed an increase in ROC AUC over individual modalities by 0.03 (\textit{p>0.221}), but only at the 12m horizon. When either DESS or TSE was used in combination with the $T_2$map, no clear and consistent differences were observed compared to just the structural MR sequence. Against the individual $T_2$map modality, the models yielded an increase by 0.02-0.04 of ROC AUC, which was, however, significant (\textit{p=0.010}) for model \textit{F5} at 36m target and insignificant (\textit{p>0.057}) elsewhere. The same models were able to marginally improve the AP scores at the 12m horizon by 0.02-0.03 (\textit{p>0.375}) over individual TSE and $T_2$maps, but not higher than the DESS model. Otherwise, no noticeable difference in AP was observed. Among the MR2 models, DESS with TSE was marginally better for 12-24m horizons in ROC AUC, while DESS with $T_2$map was more dominant at 36-48m in both metrics.

\paragraph{Fusion of multiple imaging modalities}
A combination of radiographic and single-protocol MRI images generally resulted in a performance similar to the latter, yet a few notable improvements were observed in the ROC AUC space. Namely, the \textit{F1} model (XR, DESS) showed an increase of 0.11 (\textit{p=0.039}) and 0.05 (\textit{p=0.106}) in the score at the 12m horizon compared to the individual XR and MRI DESS modalities, respectively. With the model \textit{F3} (XR, $T_2$map), the gains of 0.03 (\textit{p=0.103}) and 0.02 (\textit{p=0.177}) in ROC AUC were observed over \textit{M3} at the 48m and 96m horizons. Several performance drops were observed for the model \textit{F3} at 12m (by 0.08) and all the models \textit{F1-F3} at 24m (by 0.01-0.04) horizons. In terms of AP, the \textit{F1} model showed a marginal gain of 0.02 (\textit{p>0.238}) for 48m and 96m targets over the model \textit{M1}. The models \textit{F2} (XR, TSE) and \textit{F3} yielded rather consistent performance regression of 0.01-0.04 at all targets compared to the corresponding models \textit{M2} and \textit{M3}.

In the setting with 3 modalities (XR and two MR sequences), the scores were largely similar to the XR1MR1 models. However, both ROC AUCs and APs recovered to the level highest across the included individual modalities at 12m-36m horizons. Compared to the corresponding MR2 models, the metrics were also generally similar, with an exception being the 12m and 48-96m horizons. At the 12m target, the ROC AUCs further improved over MR2 by 0.01-0.04 (\textit{p>0.090}), which resulted in the model \textit{F7} being significantly (\textit{p=0.021}) better than the model \textit{X} and the model \textit{F8} - over \textit{X} (\textit{p=0.005}) and \textit{M1} (\textit{p=0.026}). At 48m and 96m targets, a marginal consistent gain of 0.01 over MR2 was observed in all models XR1MR2 in both metrics. Overall, the top performing model was \textit{F8} (XR, DESS, $T_2$map), yielding the highest number of statistically significant improvements over the individual clinical and imaging modalities.

\paragraph{Fusion of all imaging modalities and clinical data}
Lastly, the modalities from the best performing model \textit{F8} were combined with the clinical variables in a holistic fusion model \textit{U}. Here, the XR1MR2 architecture was extended with an additional shallow fully connected branch to embed the clinical variables (see Figure~\ref{subfig:archs_d}). The model demonstrated a performance similar or marginally lower to the one without clinical variables, namely, 0.70-0.76 in ROC AUC across the targets and 0.10~(0.02), 0.15~(0.03), 0.23~(0.03), 0.26~(0.03), and 0.55~(0.03) in AP for 12m, 24m, 36m, 48m, and 96m horizons respectively. Interestingly, the model \textit{U} was not able to achieve the highest AP at the 12m target, demonstrated previously by \textit{C3} model.

\subsection*{Performance with respect to patient sub-groups}
The performance of models on the heterogeneous patient cohorts brings rather limited interpretation capabilities and, thus, actionable insights. To explore which patients may benefit from using certain imaging modalities and predictive models, we analyzed the performance metrics sub-group-wise. Here, we selected only those subjects, for whom the labels were available at all the horizons. Next, all the subjects were assigned to one of the three groups - "no prior injury or surgery", "prior injury, but no surgery", or "prior surgery". The prevalence rates of progressors in the groups were 0.059, 0.106, and 0.067, respectively. Post-traumatic cases may show distinct imaging findings and are often considered separate phenotypes in scientific literature~\cite{anderson2011post,mobasheri2019recent}, thus, such separation. Within each of these groups, the subjects were further divided into sub-groups, based on the severity of radiographic KOA ("KLG 0-1", "KLG 2", "KLG 3") and presence of symptoms ("WOMAC 0-10", "WOMAC 10-100"). Within each sub-group, we calculated the performance metrics by averaging them over all the horizons. For AP, to account for different prevalences across the targets, the metric was calibrated before averaging to a fixed prevalence of 0.15~\cite{siblini2020master}. The models compared included the individual modalities -- clinical, X-ray, and DESS MRI --, as well as the top-ranked multi-modal fusion model. The latter was selected via a multi-objective ranking procedure over all horizons and both performance metrics (see the details in Methods).

\begin{figure}[ht]
\centering\subfloat[\label{subfig:groupw_a}]{\includegraphics[clip,trim=0cm 0.3cm 0cm 0.3cm,width=.33\linewidth]{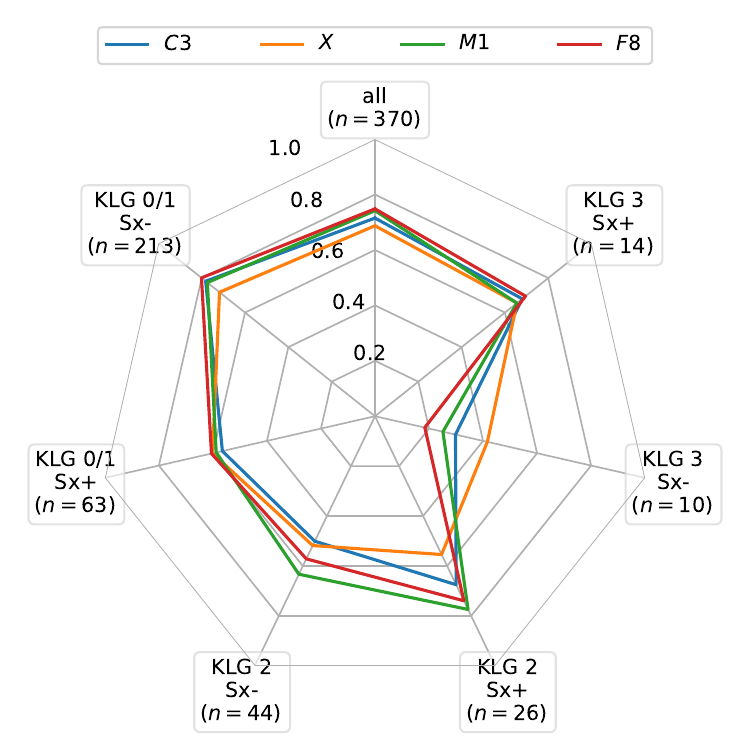}}
\hfil
\centering\subfloat[\label{subfig:groupw_b}]{\includegraphics[clip,trim=0cm 0.3cm 0cm 0.3cm,width=.33\linewidth]{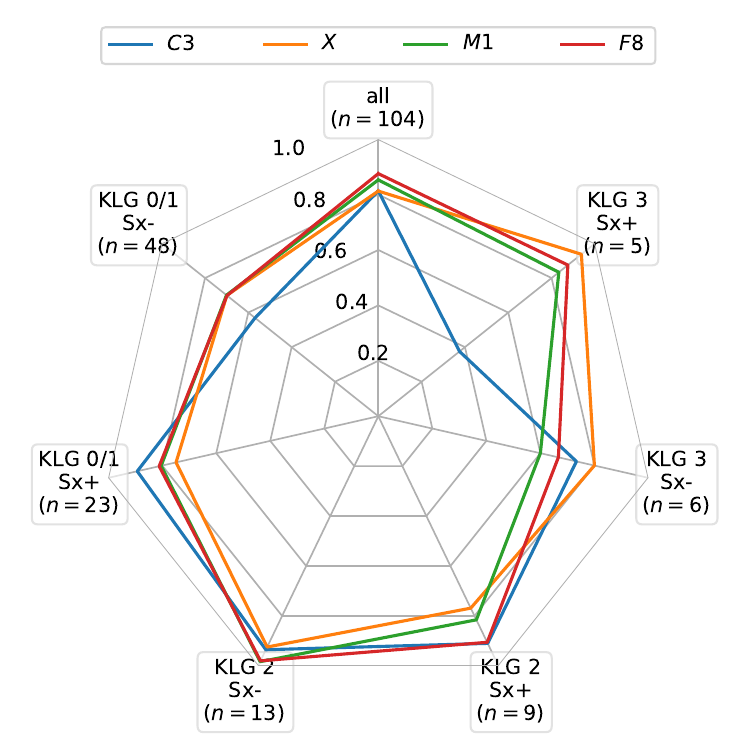}}
\hfil
\centering\subfloat[\label{subfig:groupw_c}]{\includegraphics[clip,trim=0cm 0.3cm 0cm 0.3cm,width=.33\linewidth]{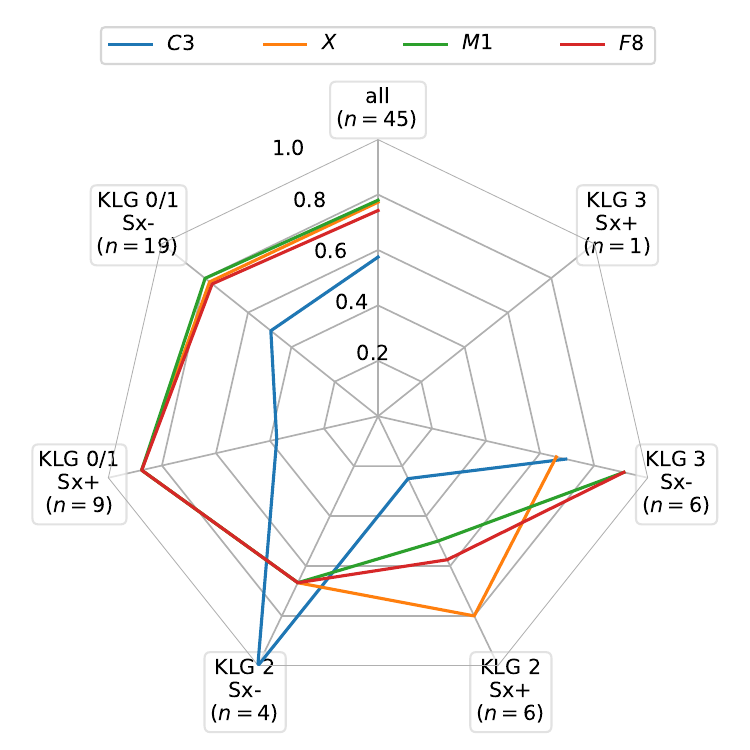}}
\vfil
\centering\subfloat[\label{subfig:groupw_d}]{\includegraphics[clip,trim=0cm 0.3cm 0cm 1.3cm,width=.33\linewidth]{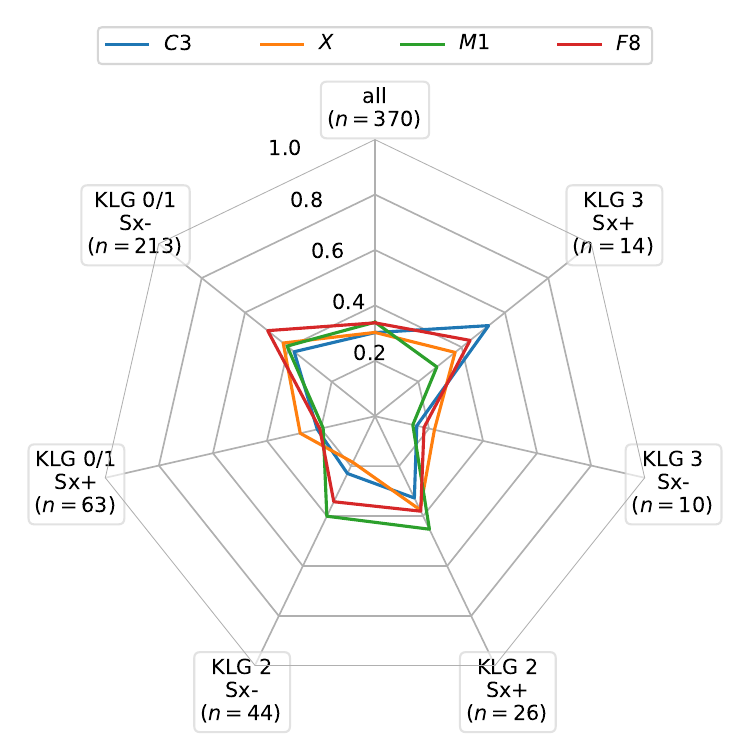}}
\hfil
\centering\subfloat[\label{subfig:groupw_e}]{\includegraphics[clip,trim=0cm 0.3cm 0cm 1.3cm,width=.33\linewidth]{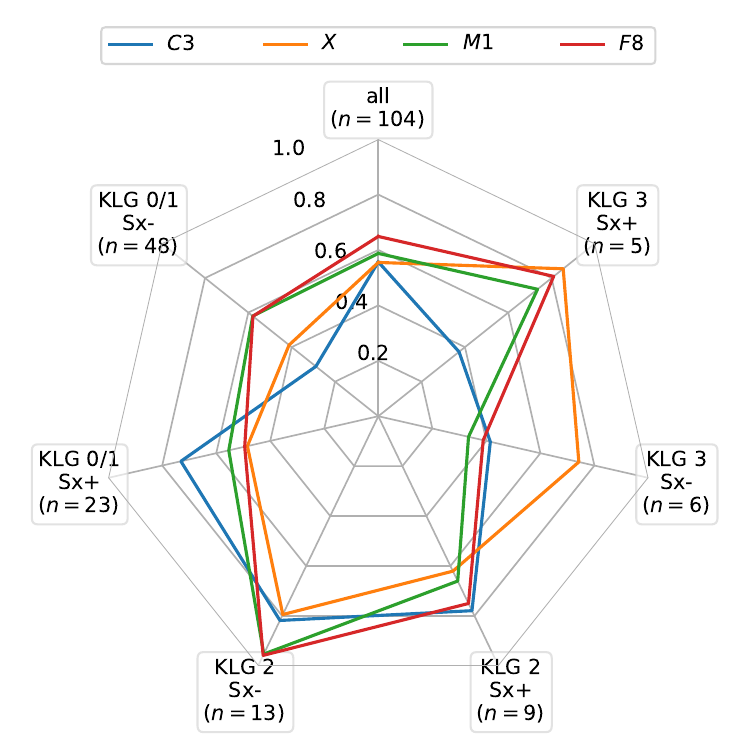}}
\hfil
\centering\subfloat[\label{subfig:groupw_f}]{\includegraphics[clip,trim=0cm 0.3cm 0cm 1.3cm,width=.33\linewidth]{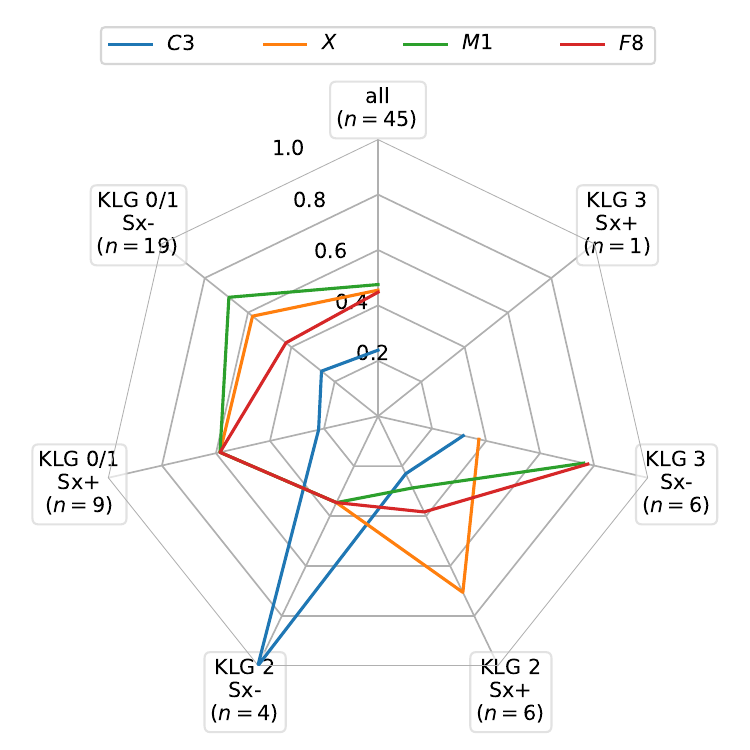}}
\caption{Performance of the selected models in subject subgroups averaged over all the horizons. The subjects are stratified by their trauma and intervention history - \textit{"no prior injury or surgery"} (\Cref{subfig:groupw_a,subfig:groupw_d}), \textit{"prior injury, but no surgery"} (\Cref{subfig:groupw_b,subfig:groupw_e}), \textit{"prior surgery"} (\Cref{subfig:groupw_c,subfig:groupw_f}). The plots show mean area under the ROC curve (\Cref{subfig:groupw_a,subfig:groupw_b,subfig:groupw_c}) and average precision (\Cref{subfig:groupw_d,subfig:groupw_e,subfig:groupw_f}). The precision values were calibrated\cite{siblini2020master} to the prevalence of \textit{0.15} before averaging. Each plot indicates the scores for the complete corresponding sample (\textit{all}), as well as mutually exclusive sub-groups allocated w.r.t severity of radiographic OA (\textit{KLG~0/1} vs. \textit{KLG~2} vs. \textit{KLG~3}) and symptoms (\textit{WOMAC~0-10~[Sx-]} vs. \textit{WOMAC~10-100~[Sx+]}).}
\label{figure:groupw}
\end{figure}

We first considered the "no prior injury or surgery" group. Here, the overall ROC AUCs were moderate with all the models. The highest performance (AUC=0.65-0.80) was observed in asymptomatic KLG0/1, as well as symptomatic KLG2 and KLG3 sub-groups. The X-ray model was more consistent across the sub-groups but was inferior to other models for symptomatic KLG2 subjects. All models performed poorly with the asymptomatic KLG3 sub-group (AUC<0.50), which was also the smallest one. In terms of AP, the performance was generally low (AP=0.20-0.55), showing the challenging nature of the OA progression prediction problem. MRI- (\textit{M1}) and fusion-based (\textit{F8}) models performed stronger with asymptomatic KLG0/1, all KLG2, and symptomatic KLG3 subjects.

In the "prior injury, but no surgery" group, the overall performance in ROC AUC was high-to-very-high, with \textit{M1} and \textit{F8} models showing an increase up to 0.10 over the rest (Figure~\ref{subfig:groupw_b}). Here, the imaging models showed high AUC in all the sub-groups. The models using MRI were more accurate at KLG0-2, while the XR model was slightly more accurate at KLG3. In AP, \textit{M1} and \textit{F8} were dominant in the same sub-groups as previously (Figure~\ref{subfig:groupw_e}). The model based on the clinical data showed the highest score in the symptomatic KLG0/1 sub-group and was comparable at KLG2, otherwise performing poorly. The X-ray-based model was more accurate towards severe OA stages, particularly, at KLG3. Both metrics were notably higher than in the "no prior injury or surgery" subject group, suggesting the clear added value of imaging, particularly, MRI in post-traumatic subjects.

In the "prior surgery" group analysis all the considered imaging models showed moderate-to-very-high ROC AUCs. Importantly, all the sub-groups here had very small sample sizes. The clinical model was notably inferior in performance, except for the small asymptomatic KLG2 sub-group. \textit{M1} and \textit{F8} showed performance similar to each other, with the former having much higher AP for the asymptomatic KLG0/1 sub-group. The X-ray model was more accurate in both metrics for the symptomatic KLG2 sub-group.

To summarize the findings, the performance of all the models in predicting KOA progression was consistently higher in post-traumatic and post-intervention knees. In the same groups, the imaging models showed more notable improvement over the clinical variable model, particularly, in positive predictive value. In the "no prior injury or surgery" group, the APs were poor with all models. However, the imaging with MRI provided additional value for normal and mild OA knees. Interestingly, the fusion model to a degree resembled the average performance of XR- and DESS MRI-based models. 

\subsection*{Contribution of imaging modalities in multi-modal setting}
\begin{figure}[ht]
\centering\subfloat[\label{subfig:util_a} {\textit{F1}~(XR, DESS)}]{\includegraphics[width=.245\linewidth]{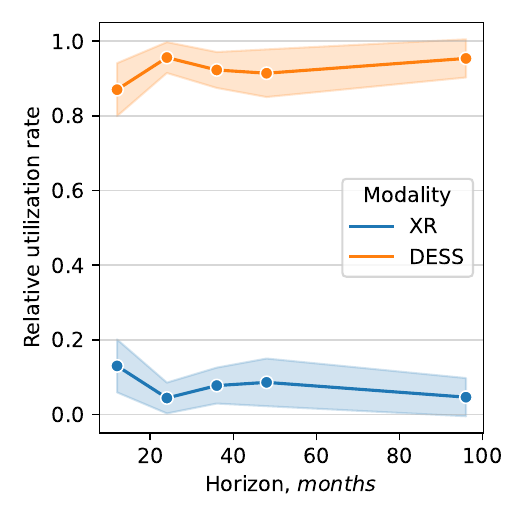}}
\hfil
\centering\subfloat[\label{subfig:util_b} {\textit{F5}~(DESS, T$_2$\text{map})}]{\includegraphics[width=.245\linewidth]{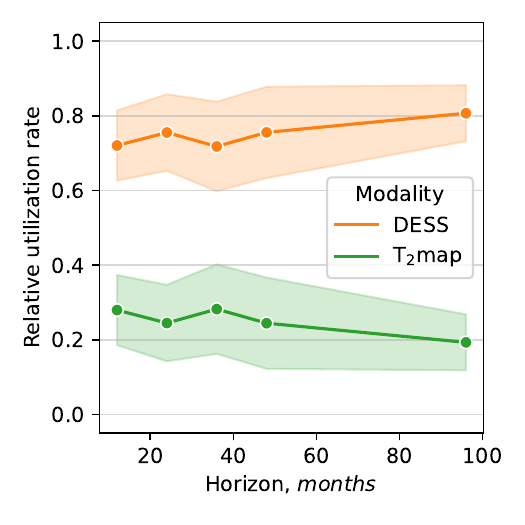}}
\hfil
\centering\subfloat[\label{subfig:util_c} {\textit{F8}~(XR, DESS, T$_2$\text{map})}]{\includegraphics[width=.245\linewidth]{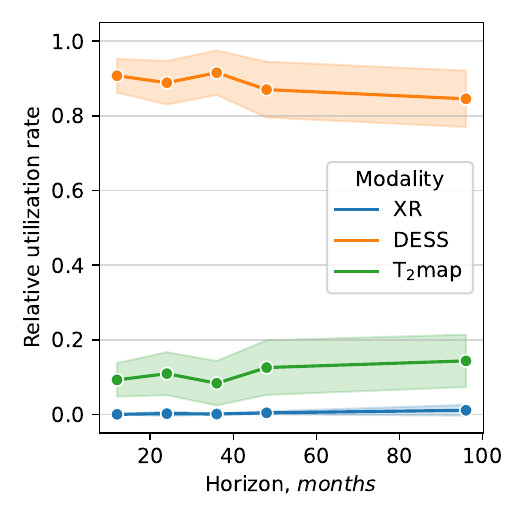}}
% \hfil
\centering\subfloat[\label{subfig:util_d} {\textit{U}~(XR, DESS, T$_2$\text{map}, clin.)}]{\includegraphics[width=.245\linewidth]{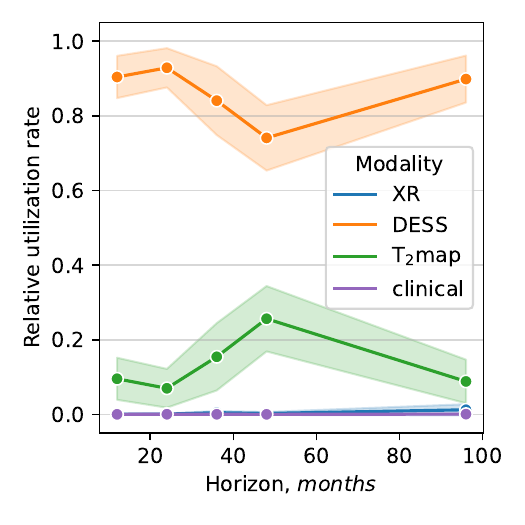}}
\caption{Relative utilization rate of individual modalities in the top performing fusion models. \textit{Horizon} represents different intervals within which the progression is considered (correspond to the prediction targets). Means (solid lines) and standard deviations (color bands) are computed over the test subset samples. The discrete horizons (12, 24, etc months) are connected via linear interpolation for visualization purposes.}
\label{figure:utilization}
\end{figure}

To understand the relative contribution of imaging modalities to the final decision in the top-performing fusion models, a model interpretation technique called "feature ablation" was employed. Here, the entire inputs corresponding to the modalities were individually masked, and the drop in the model performance was recorded. The decrements were inverted and normalized across the modalities to derive Relative Utilization Rate (RUR).
The RURs computed for the selected models are shown in Figure~\ref{figure:utilization}. In the case where radiographic and structural MRI (DESS) data were fused, the average contributions were 0.04-0.13 and 0.87-0.96, respectively, across the horizons (Figure~\ref{subfig:util_a}). This suggests that the anatomical information provided by the volumetric MRI scan is dominantly more informative in the scope of radiographic KOA progression prediction. 

When structural (DESS) and compositional ($T_2$map) MRI protocols were considered together (Figure~\ref{subfig:util_b}), the average RURs were 0.72 and 0.28 at 12m horizon and they gradually changed to 0.81 and 0.19 at 96m horizon, respectively. The reduced RUR for DESS MRI may indicate the importance of tissue compositional changes provided with $T_2$map in the scope of KOA progression, but also that certain imaging biomarkers are more easily derived from high-contrast $T_2$maps. The observed trend from 12m towards 96m horizon may indicate lower overall importance of the visualized tissue composition (particularly, cartilage) on the progression long-term. The model fusing radiographic data with two MRI protocols (Figure~\ref{subfig:archs_c}) also showed that volumetric structural data dominates other imaging sources (0.85-0.92 [DESS] versus 0.08-0.14 [$T_2$map] and <0.02 [XR]). Interestingly, the model assigned very low RUR to the XR modality. When the clinical data were additionally incorporated into the model (Figure~\ref{subfig:util_d}), it also barely showed any contribution at all the horizons (average RURs<0.01). Overall, these findings suggest that MRI-based modalities are highly informative and visualize symptomatic, post-surgical, and post-traumatic cues at the level or higher than the clinical variables and X-ray data that are relevant to radiographic KOA progression.

%%%%%%%%%%%%%%%%%%%%%%%%%%%%%%%%%%%%%%%%%%%%%%%%%%%%%%%%%%%%%%%%%%%%%%%%%%%%%%%%%%%%%%%%%%%%%%%%%%%%
\section*{Discussion}\label{sec:discussion}
In this study, we presented a multi-modal method for prediction of radiographic KOA progression and applied it to perform an exhaustive study of commonly acquired modalities in the task. Our proposed approach enables leveraging unique large-scale longitudinal cohorts, such as OAI, for studying the disease progression in broad populations.

The primary finding of our work is that the fusion of multiple widely acquired modalities, particularly, imaging, does not seem to provide significant improvement in the prediction of knee osteoarthritis progression, defined as radiographic worsening, over single modalities, both in short- and long-term horizons. It is important to note, however, that the overall best-ranked model in our experiments was based on XR, structural (DESS), and compositional ($T_2$map) MRI, suggesting that some of the subjects may still benefit from the multi-modal examination.
We have shown that $T_2$maps seem to have marginal additional value in all prediction horizons. This may be partially explained by the potentially limited association between compositional tissue properties and KOA progression defined radiographically.
Furthermore, unresolved methodological challenges, such as considerable field orientation dependence of $T_2$, might have also contributed to this finding~\cite{hanninen2021orientation,leskinen2023t2}. Importantly, we also acknowledge the fact that the studied MRI protocols, despite providing excellent contrast for major tissues, such as cartilage, bone, menisci, fat, and associated lesions, may still provide incomplete details on the knee status. The emerging imaging methods, particularly, Magnetic Resonance Fingerprinting~\cite{ma2013magnetic}, have the potential to perform holistic parametric tissue mapping and, thus, deliver a more objective view on the value of KOA MR imaging, however, they are still in the process of getting wide adoption.

Generally, all the imaging models yielded larger gains on top of the clinical data models towards longer progression horizons. This finding suggests that the role of imaging biomarkers in shorter-term progression prediction is lower, and other factors, such as subject metabolic health, environmental factors, or physical activity, may be more informative than imaging. From the practical utility perspective, using structural MRI sequences led to consistent, yet non-significant improvements over the model trained on radiographic images. While, currently, MRI is a rather expensive imaging modality, recent development in low-field MRI and fast multi-parametric techniques (e.g. aforementioned MR Fingerprinting) hold great promise that MRI could eventually become an affordable tool for osteoarthritis screening. It is important to note that not all subjects may necessarily benefit from imaging. In our sub-group analysis, we observed that the performance of the predictive models was heterogeneous and, at least, depended on whether the knee was subject to trauma, intervention, or neither. This finding suggests also that post-traumatic and post-surgical subjects should be considered independently in future scale imaging studies~\cite{anderson2011post}.

In our study, we defined the OA progression as an increase in KLG score. While KLG is the most established and widespread grading scheme for OA, it naturally lacks sensitivity to fine-grained joint changes that are not reflected directly or indirectly in radiographic images. Further works could explore more comprehensive grading schemes for the task, such as MRI-based MOAKS~\cite{hunter2011evolution}. However, this comes with a challenge -- how to define the common progression trajectory from multivariate scoring data~\cite{runhaar2014define}. Here, already existing considerations on OA phenotypes can be used~\cite{mobasheri2019recent,lee2022personalized}, however, they still require a thorough validation. Accordingly, the development of new OA surrogates in a data-driven manner could be an exciting area for future research.

In this work, we aimed to clarify the value of imaging modalities in the prediction of radiographic OA progression within multiple horizons. When targeted for downstream clinical use, DL could be used within other established domain-specific frameworks, such as time-to-event~\cite{lee2022artificial} or disease trajectory forecasting~\cite{halilaj2018modeling,lee2021ensemble,nguyen2022climat}.
Next, we used the data from a single observation point to produce the predictions. With the high-dimensional imaging data, it may be beneficial for the predictive model not only to rely on the joint anatomy but also on the rate of change derived from several successive exams of an individual. While this approach has been proven feasible for individual tissues~\cite{caliva2022surface}, processing multiple complete 3D knee scans could be an expensive computational problem, and the development of new methods is still needed.

Overall, computational and data efficiency is an important issue in multi-modal data fusion. Having larger sample sizes would likely be beneficial both for improving the performance and robustness of our models. Alternatively, modifications to the fusion model architecture can be done to reduce the number of parameters, e.g. via alternating or factorized attention in transformers~\cite{chu2021twins,arnab2021vivit}. Further works could also investigate emerging foundation models for medical imaging~\cite{chambon2022adapting,moor2023foundation}, which are to provide generic medical-specific visual features, thus, notably reducing data demand.
Finally, as previously discussed, other modalities/factors could be studied in the problem, particularly, subject lifestyle, physical activity, and metabolic biomarkers.

We interpreted the relative contribution of imaging modalities within the fusion models and observed that the structural DESS MRI was dominant across all the horizons. Such protocol certainly provides higher information and a more comprehensive view on the knee joint status. A recent study~\cite{wu2022characterizing} suggested that DL-based models are prone to greedy learning, at least, in multi-view fusion scenarios, which practically leads to unequal optimization rates across the modality branches. While the effect of this finding on the performance shown by the authors was rather small, its magnitude with diverse modalities of different shapes needs further investigation. Furthermore, the fusion of modalities may be orchestrated in a more clinically meaningful way, where using highly accessible data (e.g. clinical variables or XR images) is prioritized during the model training. Given the scope of this study, we intentionally focused on high-level model interpretability. We acknowledge that finer methods for feature attribution exist and have been applied in the KOA studies~\cite{tiulpin2019multimodal,tolpadi2020deep}, yet their generalization and applicability to multi-modal imaging settings may not be straightforward~\cite{jin2021one,jin2023guidelines}.

We hope that the findings from our study, along with the publicly released source code, will facilitate further advances in the data-driven development of knee OA progression surrogates, efficient OA progression prediction models, and clinical guidelines for OA screening.

%%%%%%%%%%%%%%%%%%%%%%%%%%%%%%%%%%%%%%%%%%%%%%%%%%%%%%%%%%%%%%%%%%%%%%%%%%%%%%%%%%%%%%%%%%%%%%%%%%%%
\section*{Methods}\label{sec:methods}

\begin{figure}[ht]
\centering
\includegraphics[height=0.91\textheight,keepaspectratio]{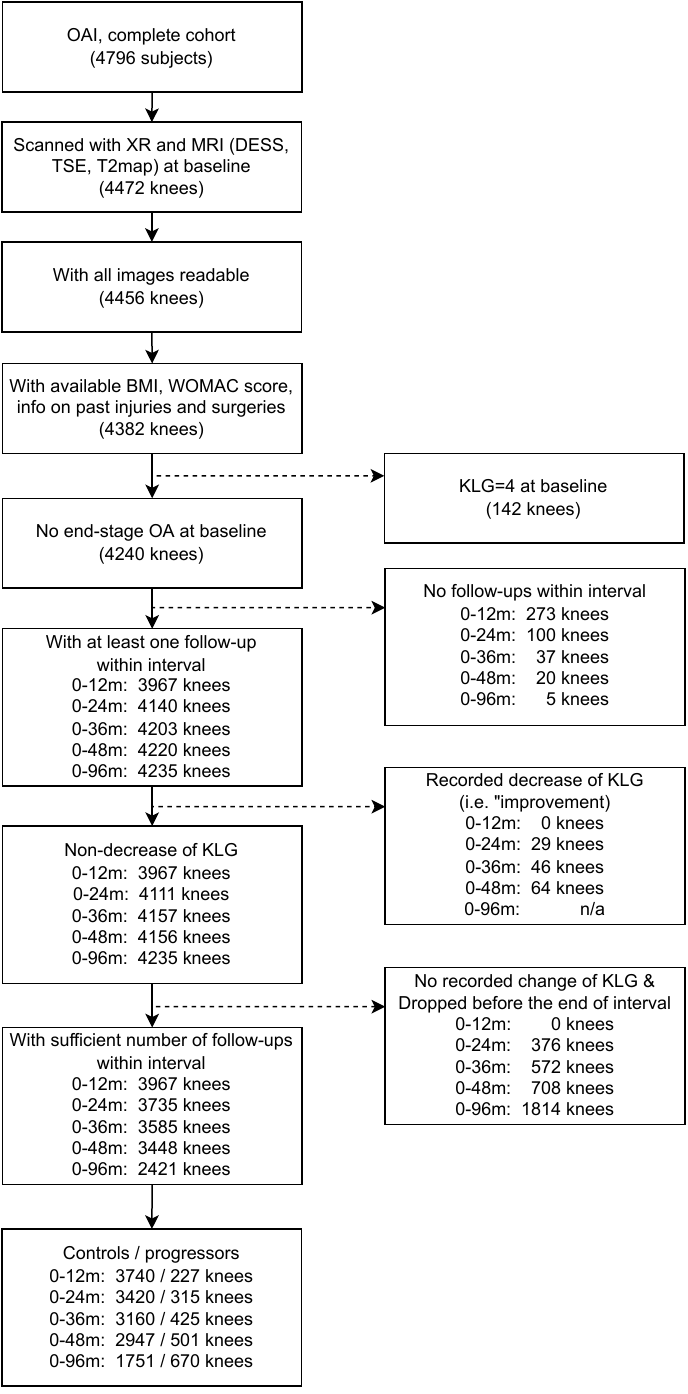}
\caption{Sample selection procedure implemented in the study. To maximize the number of samples and keep the diversity of the cohort high, the exclusion criteria mainly addressed missing variables and images, and insufficient follow-ups affecting derivation of the progression targets. For $\textit{0-96m}$ dataset, \textit{"Recorded decrease of KLG"} was not applied as an exclusion criteria, so to allow direct comparison with the study of Tiulpin et al.\cite{tiulpin2019multimodal}.}
\label{figure:selection}
\end{figure}

\paragraph*{Sample selection}
The data from The Osteoarthritis Initiative (OAI, \url{https://nda.nih.gov/oai/}) -- a multi-center longitudinal osteoarthritis study -- was used in this work. We derived five datasets from the baseline visit of OAI, one per studied progression horizon -- 12, 24, 36, 48, and 96 months (see Table~\ref{table:demogr}). All the selected subjects had demographic and clinical variables recorded, and their studied knees were imaged with posteroanterior bilateral X-ray and underwent comprehensive MRI examination (3T Siemens MAGNETOM Trio, quadrature T/R knee coils). Obtained X-ray images were weight-bearing and imaged in fixed flexion using a SynaFlexer positioning frame (CCBR-SYNARC, San Francisco, CA). The MRI exam included, among others, 3 MRI sequences - sagittal 3D dual-echo steady state (DESS, voxel $0.37\times0.37\times0.7mm$, matrix $384\times384$, 160 slices, FOV $140mm$, TR $16.3ms$, TE $4.7ms$, flip angle $25^\circ$), coronal intermediate-weighted turbo spin-echo (TSE, voxel $0.37\times0.37\times3.0mm$, matrix $384\times384$, 31 slices, FOV $140mm$, TR $3.0ms$, TE $29ms$, flip angle $180^\circ$), and sagittal multi-slice multi-echo $T_2$ mapping ($T_2$map, voxel $0.31\times0.31\times3.0mm$, matrix $384\times384$, 27 slices, FOV $120mm$, TR $2.7s$, TE $10-70ms$). Since $T_2$maps were only acquired for right knees, only one knee per subject was included. The knees within each dataset were marked as "progressor" if an increase in KLG was recorded during the respective follow-up period, and as "non-progressors" if there was no change in KLG between the baseline and the end of the interval. A small number of knees that showed an improvement in KLG during the interval was excluded. The complete sample selection procedure is provided in detail in Figure~\ref{figure:selection}.

\paragraph*{Clinical variables}
Widely acquired demographic variables, history of past injuries and past surgeries, symptomatic and knee function score - Western Ontario and McMaster Universities Arthritis Index (WOMAC), and radiographic OA severity - Kellgren-Lawrence grade (KLG) were considered. The continuous variables -- age, history of past injuries and past surgeries, body mass index (BMI), and WOMAC total score -- were standardized to zero mean and unit variance. The categorical variables -- sex, KLG, history of past injuries, and past surgeries -- were transformed using one-hot encoding.

\paragraph*{X-ray images}
The ROIs were extracted from the bilateral posteroanterior X-ray images. For that, the DL-based tool KNEEL \cite{tiulpin2019kneel} was used, which was previously developed and validated on the OAI data. The tool localized a set of bone surface landmarks in the femorotibial joint area. The landmarks were aggregated to derive the location of the knee joint center. The ROIs of $140\times140~mm$ were cropped around the knee centers. The obtained patches were resampled to an isotropic pixel spacing of $0.195\times0.195~mm^2$.

After extraction of the knee ROIs, they were further cropped to the central patches of $700\times700~pixels$. Before feeding the data into the model, the patches were first standardized in intensity to $[0; 1]$ range, underwent data augmentation (for the training samples only), and finally standardized to zero mean and unit range. Data augmentation included cropping to a random $700\times700~pixels$ patch instead of the center one, random rotation within $[-15, 15]$ degree range, and random gamma correction with $\gamma$ from the range $[0.0; 2.0]$. Lastly, the patches were downsampled using bilinear interpolation to $350\times350~pixels$ (pixel spacing of $0.390\times0.390~mm^2$).

\paragraph*{MR images}
One of the aims of MR image preprocessing was to reduce the storage and memory demand while maintaining the ROI size and the visual quality of the samples. In DESS and TSE sequence data, the 3 least significant bits were truncated, resulting in 8 significant bits for DESS and 9 bits for TSE. Subsequently, the images were clipped in intensity to $[0.0; 99.9]$ percentile range scan-wise.
For all the sequences, to exclude the image registration artifacts, we cut the slice edges of 16 voxels.

$T_2$maps were derived from the multi-slice multi-echo images via exponential fitting. On average, the OAI $T_2$ mapping acquisition protocol yielded 27 slices over 7 echo times. We used the $T_2$ relaxation monoexponential model (Equation~\ref{equat:t2}) and optimized both $I_{0}$ and $T_2$ parameters voxel-wise using the available raw image intensities $I_{TE_{i}}$ and the corresponding echo times $TE_{i}$. All the available echoes were used for fitting. The obtained $T_2$maps were clipped in intensity to $[0; 100]~ms$ range. Since the $T_2$ mapping protocol in the OAI is optimized for cartilage tissues, this helped to ensure that unreliable T2 values, which corresponded mainly to bone and fat pads, are excluded~\cite{pai2008comparative}. An example of the resulting $T_2$map is shown in Figure~\ref{figure:pipeline}.
\begin{equation}\label{equat:t2}
    I_{TE_{i}} = I_{0} \times exp(- TE_{i}/T_2)
\end{equation}

In the next step, the images were cropped to the central area of [320, 320, 128] voxels for DESS, [320, 320, 32] for TSE, and [320, 320, 25] for $T_2$maps, where the first two dimensions correspond to the number of voxel rows and voxel columns in-slice, respectively, and the last dimension corresponds to the number of slices. Similarly to the radiographic data, the images were then transformed to $[0; 1]$ intensity range, augmented, and standardized to zero mean and unit range. Data augmentation started with random cropping to the aforementioned dimensions, in-slice rotation (random degree from $[-15, 15]$ range), and gamma correction (random $\gamma$ from $[0.0; 2.0]$ range). The gamma correction was not applied to the $T_2$maps. Finally, the images were downsampled using trilinear interpolation to [160, 160, 64] voxels for DESS, [160, 160, 32] for TSE, and [160, 160, 25] for $T_2$maps.

\paragraph*{Clinical data baselines}
An independent logistic regression model was constructed for each target and each considered set of clinical variables (scikit-learn, version 0.24.2~\cite{pedregosa2011scikitlearn}). In every setting, 5-fold cross-validation was used on the development data subset to find the best hyper-parameter -- whether to use balanced class-weighting. Subsequently, 5 models were optimized using average precision scoring on the training data and evaluated on the testing subset. The ensemble predictions were derived by averaging softmax outputs across the folds.

\paragraph*{Imaging model architectures}
The imaging model architectures varied depending on the considered set of modalities while following the same design principles. A schematic description of the architectures is shown in Figure~\ref{figure:archs}, with more details provided in Section~\ref{subsec:sup_architectures} and the accompanying source code (PyTorch, version 1.8.2~\cite{paszke2019pytorch}). For radiographic data processing, we reimplemented the previously validated model~\cite{tiulpin2019multimodal} based on pre-trained {ResNeXt-50\textunderscore32x4d} CNN (see Figure~\ref{subfig:archs_a}). For individual MRI sequences, the models comprised a shared pre-trained ResNet-50 CNN to extract slice-wise image descriptors, followed by a Transformer module to aggregate the representations across slices. Such design was previously shown to achieve higher performance compared to purely CNN-based models~\cite{panfilov2022predicting}, while also providing pre-training capability that is challenging to obtain with pure Transformers and moderate sample size. For the fusion of two modalities -- XR1MR1 and MR2, an overall similar design was used. Here, two independent CNNs were used for each of the modalities, and their outputs were concatenated before the Transformer to allow for cross-modal fusion (see Figure~\ref{subfig:archs_c}). Lastly, in the fusion of three-to-four modalities, the MRI-related branches of the model had their independent mid-level Transformers to embed the features into common latent space before combining with other sources (Figure~\ref{subfig:archs_d}). The models with clinical data input had a shallow Fully-Connected network to transform the variables before fusion. A Transformer module was used on top of concatenated multi-modal embeddings, as previously.

All the described models were trained in 5-fold cross-validation, where the splits were done maintaining the consistent distribution of target labels. The training was run until convergence with a computational budget of 60 epochs. Adam optimizer~\cite{kingma2014adam} was used with weight decay of 1e-4 and learning rate warmup (from 1e-5 to 1e-4) over 5 initial training epochs. To address the effects of severe class imbalance, Focal loss ($\gamma=2.0$) was used along with an oversampling of the minority class. The best model within each fold was chosen based on the highest average precision score at validation. The batch size was 16 for the models with at least two MRI modalities, and 32 otherwise. Hardware-wise, a computational node with 4 NVIDIA A100 GPU was used for model training, and a PC with 2 NVIDIA 2080 Ti was used for evaluation and subsequent analysis. The single model training time (i.e. one fold) for the highest sample size 0-12m target varied from 0.5 (XR) to 6.5 hours (fusion of 4 modalities).

\paragraph*{Evaluation and model comparison}
For each prediction target, the corresponding models were scored with ROC AUC and AP on the hold-out data. The mean and the standard error of each metric were estimated using bootstrapping (iter=1000) stratified by the target label. The statistical significance of improvements was assessed in two scenarios -- (1) single-modality imaging models against the best clinical model, (2) fusion models against the clinical, XR, or DESS MRI models. For this, one-sided paired permutation testing (iter=1000, SciPy, version 1.9.3~\cite{virtanen2020scipy}) was used.

For the subsequent analysis, the "best overall" multi-modal fusion setting $s^{*}$ was selected using a multi-objective ranking procedure:
\begin{equation}
    s^{*} = \underset{s\in S}{\mathrm{argmin}} \Big(\sum_{f\in\{ROC~AUC,AP\}} \sum_{t\in\{12, ..., 96\}} rank(\bar{f}(s_t))\Big), ~~S=\{F1,...,F9,U\}\\
\end{equation}
Here, every fusion setting \textit{s} was ranked from 1 to 10 (best to worst, respectively) for each target \textit{t} and in each metric independently by the mean metric value \textit{$\bar{f}$}. Then, the ranks were summed, and the model with the highest total rank was chosen.

In subgroup analysis, average model performance across different targets was derived. Since the prevalence of progressors is different for different targets, which prohibits direct averaging, instead of standard AP we used its calibrated version~\cite{siblini2020master}. Here, the scores within subgroups were calculated for target prevalence of 0.15, and only then averaged. ROC AUC scores were used unchanged. Symptomatic and non-symptomatic patient subgroups were defined based on the WOMAC total score. Clinical interpretation of WOMAC score is still rather non-standardized~\cite{woolacott2012use,womac2016summary}. We used a threshold value of 10 on a total score 0-96 scale, which is an estimate of the minimal clinically important difference~\cite{williams2012comparison,womac2016summary}.

The importance of individual modalities in the multi-modal fusion settings was estimated using the feature ablation method (Captum, version 0.5.0, Facebook Open Source~\cite{kokhlikyan2020captum}). Here, the unimodal inputs were replaced with the mean values one-by-one and degradation of the model performance was recorded for each sample. The values were normalized and averaged across the testing subset, which resulted in Relative Utilization Rates.

%%%%%%%%%%%%%%%%%%%%%%%%%%%%%%%%%%%%%%%%%%%%%%%%%%%%%%%%%%%%%%%%%%%%%%%%%%%%%%%%%%%%%%%%%%%%%%%%%%%%
\newpage
\bibliography{literature}

%%%%%%%%%%%%%%%%%%%%%%%%%%%%%%%%%%%%%%%%%%%%%%%%%%%%%%%%%%%%%%%%%%%%%%%%%%%%%%%%%%%%%%%%%%%%%%%%%%%%
\section*{Acknowledgements}\label{sec:acknowledgments}
\anonymized{The authors acknowledge the following funding sources: strategic funding of Infotech Institute, University of Oulu; 6GESS Profiling Research Programme (Academy of Finland project 336449); Orion Research Foundation, Finland. CSC – IT Center for Science, Finland is kindly acknowledged for providing the generous computational resources, which made the study possible. Khanh Nguyen is acknowledged for preprocessing the radiographic images. We also thank Dr. Valentina Pedoia for an insightful discussion on the topic of the study.}

The OAI is a public-private partnership comprised of five contracts (N01-AR-2-2258; N01-AR-2-2259; N01-AR-2-2260; N01-AR-2-2261; N01-AR-2-2262) funded by the National Institutes of Health, a branch of the Department of Health and Human Services, and conducted by the OAI Study Investigators. Private funding partners include Merck Research Laboratories; Novartis Pharmaceuticals Corporation, GlaxoSmithKline; and Pfizer, Inc. Private sector funding for the OAI is managed by the Foundation for the National Institutes of Health. This manuscript was prepared using an OAI public use data set and does not necessarily reflect the opinions or views of the OAI investigators, the NIH, or the private funding partners.

%%%%%%%%%%%%%%%%%%%%%%%%%%%%%%%%%%%%%%%%%%%%%%%%%%%%%%%%%%%%%%%%%%%%%%%%%%%%%%%%%%%%%%%%%%%%%%%%%%%%
\section*{CRediT author statement}\label{sec:author}

\anonymized{\textbf{Egor Panfilov}: Methodology; Software; Formal Analysis; Investigation; Data Curation; Writing-Original Draft; Visualization. \textbf{Miika T. Nieminen}: Project Administration; Funding Acquisition. \textbf{Simo Saarakkala}: Project Administration; Funding Acquisition. \textbf{Aleksei Tiulpin}: Methodology; Data Curation; Supervision; Funding Acquisition. \textbf{All authors}: Conceptualization; Writing-Review and Editing; Final Approval.}

%%%%%%%%%%%%%%%%%%%%%%%%%%%%%%%%%%%%%%%%%%%%%%%%%%%%%%%%%%%%%%%%%%%%%%%%%%%%%%%%%%%%%%%%%%%%%%%%%%%%
\section*{Additional information}

\subsection*{Data and code availability statement}
The data used in the study is derived from the publicly available Osteoarthritis Initiative database (\href{https://nda.nih.gov/oai/}{https://nda.nih.gov/oai/}). The source code of sample selection and subset allocation procedures, all the developed methods, and the performed analysis are made available at \anonymized{\href{https://github.com/Oulu-IMEDS/OAProgressionMMF}{https://github.com/Oulu-IMEDS/OAProgressionMMF}}.

\subsection*{Competing interests}
The authors declare no competing interests in relation to the present work.

%%%%%%%%%%%%%%%%%%%%%%%%%%%%%%%%%%%%%%%%%%%%%%%%%%%%%%%%%%%%%%%%%%%%%%%%%%%%%%%%%%%%%%%%%%%%%%%%%%%%
\clearpage
\section*{\textbf{SUPPLEMENTAL MATERIALS}}
\beginsupplement

\subsection{Architectures of the models}\label{subsec:sup_architectures}
\begin{figure}[ht]
\centering\subfloat[\label{subfig:archs_a}]{\includegraphics[height=1.9cm]{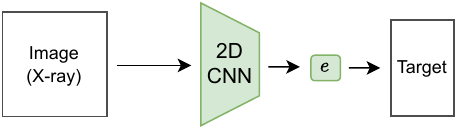}}
\vfil
\subfloat[\label{subfig:archs_b}]{\includegraphics[height=2.25cm]{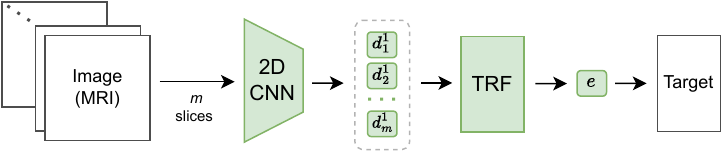}}
\vfil
\subfloat[\label{subfig:archs_c}]{\includegraphics[height=4.4cm]{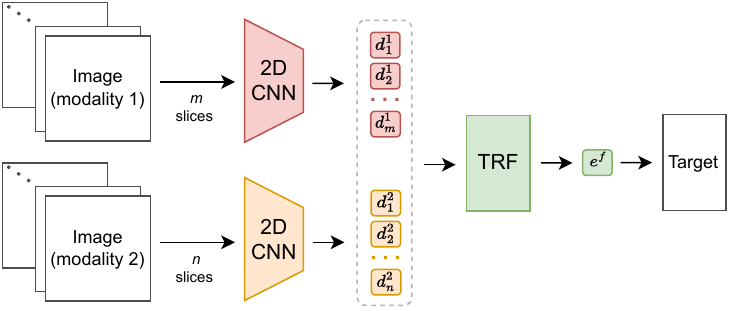}}
\vfil
\subfloat[\label{subfig:archs_d}]{\includegraphics[height=4.9cm]{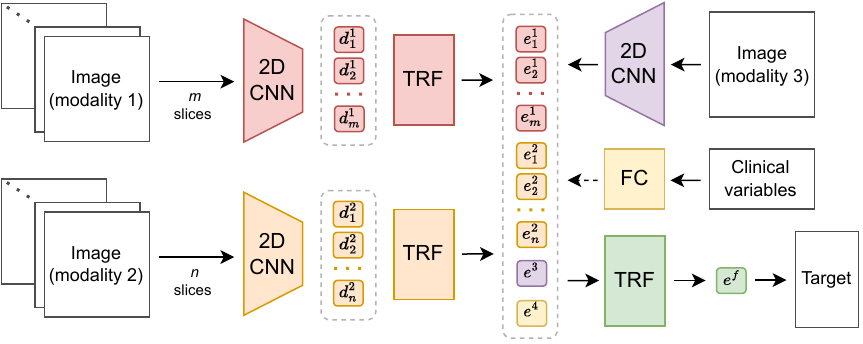}}
\caption{Neural network model architectures studied in this work. The Sub-figures correspond to the architectures from the main text as follows: \ref{subfig:archs_a} - XR1, \ref{subfig:archs_b} - MR1, \ref{subfig:archs_c} - XR1MR1 and MR2, \ref{subfig:archs_d} - XR1MR2 and XR1MR2C1. Sub-figure \ref{subfig:archs_a} comprises a conventional 2D convolutional neural network (CNN). In \Cref{subfig:archs_b,subfig:archs_c,subfig:archs_d}, the slice-level descriptors $d_{s}^{i}$ ($i$ - modality index, $s$ - slice index) are extracted with independent CNNs. The descriptors are then embedded into a common latent space $e_{s}^{i}$ with Transformers (TRFs), fused into a single representation vector $e^{f}$, and eventually associated with $target$. If incorporated, clinical variable embeddings $e^4$ are constructed using a shallow Fully Connected (FC) network.}
\label{figure:archs}
\end{figure}

The exact implementations of all the studied models can be found in the accompanying source code \anonymized{\href{https://github.com/Oulu-IMEDS/OAProgressionMMF}{[link]}}. Here, we provide only a brief overview of the architectures (see Figure~\ref{figure:archs}) along with the most important aspects. All the models were constructed by combining CNN and Transformer modules. The CNNs were ResNet-50 pre-trained on ImageNet, with the exception of the XR model, where previously developed ResNeXt-50\textunderscore32x4d was used. The latter model has previously shown stronger performance in a similar task~\cite{tiulpin2019multimodal}. The prepared images (i.e. slices) were transformed into descriptor vectors of 2048 elements by the CNNs. Next, a sequence of descriptors was passed through a Transformer (4 levels, 8 attention heads) to obtain an output of the same shape. If the Transformer was concluding the model (Figures~\ref{subfig:archs_b} and \ref{subfig:archs_c}, green in Figure~\ref{subfig:archs_d}), a fully connected network with 1 hidden layer of 2048 neurons was used to map Transformers' output to the binary target. Otherwise, the complete output state of the Transformer was propagated further. In the architecture with clinical variables, they were concatenated into a single vector and transformed to the common embedding vector of 2048 elements using a fully connected network with one layer. Dropout with the rate of 0.1 was extensively used throughout every architecture.

In the experiments reported in the article, the CNN outputs were taken after the Global Average Pooling layer. We also experimented with using non-pooled representations but did not observe any consistent improvements. For the multi-sequence MRI fusion (Figure~\ref{subfig:archs_c}), we additionally investigated the setting with a cascade of Transformers (as in Figure~\ref{subfig:archs_d}), which resulted in similar scores yet higher computational demand. Lastly, for the holistic fusion model (Figure~\ref{subfig:archs_d}) we tried mixing in the modalities one at a time, starting with XR images. This lead to generally lower performance than the one obtained with the reported architecture.

\begin{table}[ht]
\centering
\caption{Performance of the imaging fusion models. Models F1-F3 combine X-ray with single MRI sequence, F4-F6 - two MRI sequences, F7-F9 - X-ray with two MRI sequences. More details on the architecture of the fusion models are provided in Section~\ref{subsec:sup_architectures}. The values show average precision (AP) and area under the ROC curve (ROC AUC) along with the standard errors.}
\label{table:fusion}
\begin{tabular}{c|l|c|rrrrr}
\toprule
 & \multicolumn{1}{c|}{\textbf{}} & \textbf{Target $>$} & \multicolumn{1}{c|}{\textbf{12m}} & \multicolumn{1}{c|}{\textbf{24m}} & \multicolumn{1}{c|}{\textbf{36m}} & \multicolumn{1}{c|}{\textbf{48m}} & \multicolumn{1}{c}{\textbf{96m}} \\ \midrule
\textbf{Model} & \multicolumn{1}{c|}{\textbf{Data}} & \textbf{Arch} & \multicolumn{5}{c}{\textbf{ROC AUC @ target}} \\
\midrule
F1 & XR, DESS &  & \multicolumn{1}{r|}{0.76$_{\pm 0.03}^{\texttt{(x)}}$} & \multicolumn{1}{r|}{0.72$_{\pm 0.04}$} & \multicolumn{1}{r|}{0.70$_{\pm 0.03}$} & \multicolumn{1}{r|}{0.74$_{\pm 0.03}^{\texttt{(c)}}$} & 0.77$_{\pm 0.02}^{\texttt{(cx)}}$ \\
F2 & XR, TSE & XR1MR1 & \multicolumn{1}{r|}{0.71$_{\pm 0.04}$} & \multicolumn{1}{r|}{0.71$_{\pm 0.03}$} & \multicolumn{1}{r|}{0.68$_{\pm 0.03}$} & \multicolumn{1}{r|}{0.69$_{\pm 0.03}$} & 0.77$_{\pm 0.02}^{\texttt{(cx)}}$ \\
F3 & XR, T$_2$\text{map} &  & \multicolumn{1}{r|}{0.61$_{\pm 0.04}$} & \multicolumn{1}{r|}{0.68$_{\pm 0.04}$} & \multicolumn{1}{r|}{0.66$_{\pm 0.03}$} & \multicolumn{1}{r|}{0.71$_{\pm 0.03}$} & 0.76$_{\pm 0.02}^{\texttt{(cx)}}$ \\
\midrule
F4 & DESS, TSE &  & \multicolumn{1}{r|}{0.74$_{\pm 0.03}$} & \multicolumn{1}{r|}{0.74$_{\pm 0.03}$} & \multicolumn{1}{r|}{0.70$_{\pm 0.03}$} & \multicolumn{1}{r|}{0.71$_{\pm 0.03}$} & 0.76$_{\pm 0.02}^{\texttt{(cx)}}$ \\
F5 & DESS, T$_2$\text{map} & MR2 & \multicolumn{1}{r|}{0.72$_{\pm 0.04}$} & \multicolumn{1}{r|}{0.74$_{\pm 0.03}$} & \multicolumn{1}{r|}{0.72$_{\pm 0.03}^{\texttt{(c)}}$} & \multicolumn{1}{r|}{0.73$_{\pm 0.03}^{\texttt{(c)}}$} & 0.76$_{\pm 0.02}^{\texttt{(c)}}$ \\
F6 & TSE, T$_2$\text{map} &  & \multicolumn{1}{r|}{0.71$_{\pm 0.04}$} & \multicolumn{1}{r|}{0.73$_{\pm 0.03}$} & \multicolumn{1}{r|}{0.71$_{\pm 0.03}$} & \multicolumn{1}{r|}{0.72$_{\pm 0.03}$} & 0.75$_{\pm 0.02}^{\texttt{(c)}}$ \\
\midrule
F7 & XR, DESS, TSE &  & \multicolumn{1}{r|}{0.75$_{\pm 0.04}^{\texttt{(x)}}$} & \multicolumn{1}{r|}{0.73$_{\pm 0.04}$} & \multicolumn{1}{r|}{0.70$_{\pm 0.03}$} & \multicolumn{1}{r|}{0.73$_{\pm 0.03}^{\texttt{(c)}}$} & 0.77$_{\pm 0.02}^{\texttt{(cx)}}$ \\
F8 & XR, DESS, T$_2$\text{map} & XR1MR2 & \multicolumn{1}{r|}{0.76$_{\pm 0.04}^{\texttt{(xm)}}$} & \multicolumn{1}{r|}{0.75$_{\pm 0.03}^{\texttt{(c)}}$} & \multicolumn{1}{r|}{0.70$_{\pm 0.03}$} & \multicolumn{1}{r|}{0.73$_{\pm 0.03}^{\texttt{(c)}}$} & 0.77$_{\pm 0.02}^{\texttt{(cx)}}$ \\
F9 & XR, TSE, T$_2$\text{map} &  & \multicolumn{1}{r|}{0.70$_{\pm 0.04}$} & \multicolumn{1}{r|}{0.73$_{\pm 0.03}$} & \multicolumn{1}{r|}{0.69$_{\pm 0.03}$} & \multicolumn{1}{r|}{0.71$_{\pm 0.03}$} & 0.76$_{\pm 0.02}^{\texttt{(c)}}$ \\
\midrule
U & \textit{F8 \& C3 vars} & XR1MR2C1 & \multicolumn{1}{r|}{0.71$_{\pm 0.04}$} & \multicolumn{1}{r|}{0.73$_{\pm 0.03}$} & \multicolumn{1}{r|}{0.70$_{\pm 0.03}$} & \multicolumn{1}{r|}{0.72$_{\pm 0.03}^{\texttt{(c)}}$} & \multicolumn{1}{r}{0.76$_{\pm 0.02}^{\texttt{(cx)}}$} \\ 
\midrule
\textbf{Model} & \multicolumn{1}{c|}{\textbf{Data}} & \textbf{Arch} & \multicolumn{5}{c}{\textbf{AP @ target}} \\
\midrule
F1 & XR, DESS &  & \multicolumn{1}{r|}{0.11$_{\pm 0.03}$} & \multicolumn{1}{r|}{0.15$_{\pm 0.03}$} & \multicolumn{1}{r|}{0.23$_{\pm 0.03}$} & \multicolumn{1}{r|}{0.29$_{\pm 0.04}^{\texttt{(c)}}$} & 0.56$_{\pm 0.03}^{\texttt{(cx)}}$ \\
F2 & XR, TSE & XR1MR1 & \multicolumn{1}{r|}{0.10$_{\pm 0.03}$} & \multicolumn{1}{r|}{0.13$_{\pm 0.03}$} & \multicolumn{1}{r|}{0.18$_{\pm 0.02}$} & \multicolumn{1}{r|}{0.28$_{\pm 0.04}^{\texttt{(c)}}$} & 0.54$_{\pm 0.03}^{\texttt{(c)}}$ \\
F3 & XR, T$_2$\text{map} &  & \multicolumn{1}{r|}{0.09$_{\pm 0.03}$} & \multicolumn{1}{r|}{0.13$_{\pm 0.03}$} & \multicolumn{1}{r|}{0.20$_{\pm 0.03}$} & \multicolumn{1}{r|}{0.27$_{\pm 0.03}$} & 0.53$_{\pm 0.03}^{\texttt{(c)}}$ \\
\midrule
F4 & DESS, TSE &  & \multicolumn{1}{r|}{0.12$_{\pm 0.03}$} & \multicolumn{1}{r|}{0.16$_{\pm 0.03}$} & \multicolumn{1}{r|}{0.20$_{\pm 0.03}$} & \multicolumn{1}{r|}{0.27$_{\pm 0.03}$} & 0.55$_{\pm 0.03}^{\texttt{(c)}}$ \\
F5 & DESS, T$_2$\text{map} & MR2 & \multicolumn{1}{r|}{0.12$_{\pm 0.04}$} & \multicolumn{1}{r|}{0.16$_{\pm 0.03}$} & \multicolumn{1}{r|}{0.24$_{\pm 0.03}$} & \multicolumn{1}{r|}{0.28$_{\pm 0.03}$} & 0.54$_{\pm 0.03}^{\texttt{(c)}}$ \\
F6 & TSE, T$_2$\text{map} &  & \multicolumn{1}{r|}{0.13$_{\pm 0.04}$} & \multicolumn{1}{r|}{0.15$_{\pm 0.03}$} & \multicolumn{1}{r|}{0.21$_{\pm 0.03}$} & \multicolumn{1}{r|}{0.26$_{\pm 0.03}$} & 0.53$_{\pm 0.04}^{\texttt{(c)}}$ \\
\midrule
F7 & XR, DESS, TSE &  & \multicolumn{1}{r|}{0.12$_{\pm 0.04}$} & \multicolumn{1}{r|}{0.15$_{\pm 0.03}$} & \multicolumn{1}{r|}{0.24$_{\pm 0.03}$} & \multicolumn{1}{r|}{0.29$_{\pm 0.04}^{\texttt{(c)}}$} & 0.54$_{\pm 0.04}^{\texttt{(c)}}$ \\
F8 & XR, DESS, T$_2$\text{map} & XR1MR2 & \multicolumn{1}{r|}{0.13$_{\pm 0.04}$} & \multicolumn{1}{r|}{0.16$_{\pm 0.03}$} & \multicolumn{1}{r|}{0.22$_{\pm 0.03}$} & \multicolumn{1}{r|}{0.27$_{\pm 0.03}$} & 0.57$_{\pm 0.03}^{\texttt{(cx)}}$ \\
F9 & XR, TSE, T$_2$\text{map} &  & \multicolumn{1}{r|}{0.10$_{\pm 0.03}$} & \multicolumn{1}{r|}{0.13$_{\pm 0.03}$} & \multicolumn{1}{r|}{0.19$_{\pm 0.02}$} & \multicolumn{1}{r|}{0.28$_{\pm 0.03}$} & 0.54$_{\pm 0.03}^{\texttt{(c)}}$ \\ \midrule
U & \textit{F8 \& C3 vars} & XR1MR2C1 & \multicolumn{1}{r|}{0.10$_{\pm 0.02}$} & \multicolumn{1}{r|}{0.15$_{\pm 0.03}$} & \multicolumn{1}{r|}{0.23$_{\pm 0.03}$} & \multicolumn{1}{r|}{0.26$_{\pm 0.03}$} & \multicolumn{1}{r}{0.55$_{\pm 0.03}^{\texttt{(c)}}$} \\
\midrule
 & \textit{prevalence} &  & \multicolumn{1}{c|}{\textit{0.04}} & \multicolumn{1}{c|}{\textit{0.06}} & \multicolumn{1}{c|}{\textit{0.10}} & \multicolumn{1}{c|}{\textit{0.13}} & \multicolumn{1}{c}{\textit{0.25}} \\
\bottomrule
\end{tabular}
\end{table}

\end{document}